\documentclass{aa}

\usepackage{graphicx}

\usepackage{txfonts}
\usepackage[switch]{lineno}
\usepackage{verbatim}
\usepackage[colorlinks,allcolors=blue]{hyperref}
\usepackage{placeins}

\begin{document}

\title{Joint X-ray and gamma-ray analysis at the photon level: Application to MSH\,15-52}
   \author{K. Egg
          \and
          A. M. W. Mitchell
          }

\institute{Erlangen Centre for Astroparticle Physics (ECAP), Friedrich-Alexander-Universität Erlangen-Nürnberg, Nikolaus-Fiebiger-Str. 2, 91058 Erlangen, Germany\\
\email{katharina.egg@fau.de}
         }

\date{Received ?? DD, 2026; accepted ?? DD, YYYY}

  \abstract

   {Joint analysis of multi-wavelength (MWL) data at the level of photon events is challenging, but it provides a statistically preferable method of fitting the data, in contrast to fitting derived flux points. Data from X-ray and gamma-ray instruments are well-suited for a joint description, due to their common physical origin (same particle population and non-thermal processes) and incidentally also due to the high similarity of their formats. Such a description is especially desirable for objects such as pulsar wind nebulae (PWNe), diffuse sources that originate from outflows of particles that have been accelerated by pulsars. Photon emission is generated by particle interactions with the surrounding magnetic and radiation fields. The rich MWL emission from PWNe renders MWL studies essential to obtain a comprehensive picture of their physical properties.}
   {We demonstrate a joint MWL analysis of 3D X-ray and gamma-ray data at the photon event level by importing eROSITA X-ray data into the Gammapy software framework.}
   {We present our pipeline for converting eROSITA 3D (two spatial and one spectral dimension) event data into a Gammapy-readable format. We validated the approach through comparison with standard X-ray data products and analysis. Furthermore a 3D eROSITA data analysis as well as a joint fit with 3D H.E.S.S. data and Fermi 4FGL catalogue flux points was performed.}
   {We show through the analysis of data from the PWN MSH\,15-52 that 1D fits of Gammapy-extracted eROSITA spectra agree with the results of the native X-ray tool PyXspec. We demonstrate that 3D analyses on X-ray data can be conducted, with both background subtraction and background modelling. Furthermore, we show that joint analyses at event level between 3D X-ray and gamma-ray data can be conducted in Gammapy. This opens up exciting new possibilities for joint analyses of PWNe and other objects. }
   {}

   \keywords{Methods: data analysis  --
                X-rays: general -- X-rays: individuals: MSH 15-52 --
                Gamma rays: general -- Radiation mechanisms: non-thermal 
               }

   \maketitle

\section{Introduction}

Non-thermal emission generated from relativistic leptonic and/or hadronic particles is seen across the electromagnetic spectrum, from radio up to very-high energy (VHE) gamma-ray photons.
Originating from cosmic accelerators such as pulsars, these particles interact with the surrounding magnetic and radiation fields, producing synchrotron, inverse Compton, and/or pion decay radiation \citep{Hinton_2009}. In order to characterise the underlying particle population of this non-thermal emission, as well as its environment, multi-wavelength (MWL) data are required \citep{Gaensler_2006,Mitchell_2022}. This generally necessitates the usage of more than one instrument to cover an appropriately broad energy range.

The joint analysis of data acquired by more than one instrument can prove challenging, not least due to various data formats. For X-ray astronomy an important step towards resolving this issue was taken with the establishment of the universal OGIP FITS data format\footnote{\url{https://heasarc.gsfc.nasa.gov/docs/heasarc/ofwg/}} for X-ray energy spectra in the 1990s. This development was in synergy with the advent of the Xspec software, capable of instrument-independent spectral analysis \citep{Arnaud_1996}. Xspec now provides a universal framework for analysing X-ray data with the forward-folding method, which convolves a model with the detector response and compares it with the observational data and has been used for X-ray data since the 1960s (see e.g. \citet{Gorenstein_1967}).

In recent years gamma-ray astronomy has been working towards a universal data format through the definition of the gamma-astro-data-formats (GADF) data standard \citep{Deil_2017}. The even further reaching Very-high-energy Open Data Format (VODF) initiative\footnote{\url{https://vodf.readthedocs.io/en/latest/}} seeks to additionally include neutrino detectors.
Furthermore the Gammapy package \citep{Donath_2023} has become a widely used software analysis package. It is optimised not only for analysis of data from one instrument, but also towards integrating data of multiple instruments into one data format, the dataset, which allows joint multi-instrument, and even MWL, analyses. A dataset can consist of either 3D data (two spatial dimensions and one energy dimension), 1D data (only energy dimension), or flux points. However, all possible dataset formats (3D, 1D, and flux points) can be jointly fit (see e.g. \citet{Nigro_2019,Rosillo_2025}). Gammapy also employs the forward-folding technique \citep{Donath_2023}.
Given the similarities in X-ray and gamma-ray data and analysis methods, mapping one approach to the other for joint analyses is not only possible but very desirable to enable joint X-ray and gamma-ray analyses at the photon level.

A joint analysis of data at the photon event level has a number of advantages over the use of flux point data. 
Systematic errors can be included in the modelling process, by applying priors to parameters in Gammapy. 
Furthermore, especially in the soft X-ray regime where photoelectric absorption through the interstellar medium (ISM) needs to be taken into account, this absorption effect can be considered in the fit as a model component. This removes the need of extracting `unabsorbed flux points', i.e. flux points of the model corrected for the influence of absorption, from X-ray data \citep{Rosillo_2025}. Similarly the background can be included in the modelling, either derived from a source-free region or with a background model.

This approach has the major advantage of preserving the flexibility of the model and can help even complex spectral energy distribution (SED) models find their global minima without being hindered by prior assumptions made on background or absorption in flux point extraction. These prior assumptions can introduce major systematics that prevent the fit from finding the model's global minimum \citep{Rosillo_2025}.

The 3D fitting approach of simultaneous spatial and spectral fits, which Gammapy has made standard in gamma-ray astronomy, furthermore expands the classical, purely spectral, X-ray fits. By adding spatial dimensions to the data, the spatial extent of sources can be considered in the fitting process and used to separate different (even overlapping) emission components (e.g. extended sources from background). Furthermore, the exposure and effective area of the data can be treated pixel-wise, an advantage in regions with a significant exposure gradient.

In this work we demonstrate joint 3D photon event-level fitting of X-ray and gamma-ray data. We showcase this using eROSITA X-ray data and present our pipeline for converting them into a Gammapy-readable format.
eROSITA data are used due to its half-sky data availability as well as its wide field of view (FoV) of $1.03\,$deg diameter which provides an excellent match to the larger FoVs of gamma-ray instruments. We explain the steps of the conversion, validated the method against standard X-ray tools, and discuss the capabilities and advantages of this approach compared to traditional X-ray spectral fits. We also present the first 3D and joint multi-instrument fits of eROSITA X-ray data.

A concrete science case for which joint MWL data analysis can constrain physical parameters is presented by pulsar wind nebulae (PWNe). PWNe are diffuse structures around pulsars, comprised of relativistic particles, predominantly electrons and positrons, accelerated by their central pulsar. PWNe emit broadband non-thermal emission; they are especially prominent sources at VHE ($\gtrsim$100\,GeV) gamma rays, comprising the majority of all identified Galactic sources in the H.E.S.S. Galactic plane survey \citep{HGPS_2018}.
Their MWL SEDs are crucial for characterising their properties, such as their magnetic field strength \citep{Mitchell_2022}.

We validated our approach using the PWN MSH\,15-52. Its VHE gamma-ray emission was discovered by H.E.S.S. \citep{HESS_MSH_2005}, and it is also seen by Fermi-LAT in the GeV regime \citep{Fermi_MSH_2010}. At soft X-ray energies non-thermal emission has been detected by numerous missions (e.g. \citet{Seward_1983,Zhang_2008,Schoeck_2010}), making it a good candidate for an X-ray and gamma-ray MWL analysis despite its rather complicated morphology and the overlap with the neighbouring thermal remnant RCW\,89.

\section{Data and software}

\subsection{eROSITA public data}

The extended ROentgen Survey with an Imaging Telescope Array (eROSITA) telescope is the product of a German-Russian collaboration and mounted on the Spektrum-Roentgen-Gamma (SRG) satellite which orbits the L2 point of the Earth-Sun system.
eROSITA is sensitive at soft X-ray energies of $\sim 0.2 - 10 \, \mathrm{keV}$ and consists of seven telescope modules (TM1-7). Each TM essentially functions as its own individual telescope. Due to an optical light leak issue TMs 5 and 7 are often disregarded in analyses \citep{Predehl_2021,Merloni_2024}.

After a two-month long Calibration and Performance Verification (CalPV) period, eROSITA conducted four complete all-sky surveys of the X-ray sky, denoted as eROSITA All-Sky Survey (eRASS) 1 to 4 between December 2019 and February 2022\footnote{\url{https://erosita.mpe.mpg.de/erass/}}. In addition to forming a complete picture of the sky at X-ray energies, its science goals include the detection of galaxy clusters and study of X-ray point sources. The German eROSITA Consortium has access to all data taken in the western Galactic Hemisphere \citep{Merloni_2024}.

The Early Data Release\footnote{\url{https://erosita.mpe.mpg.de/edr/}} (EDR) made public the CalPV data from the German eROSITA Consortium, including a pointed exposure of PSR B1509-58, the pulsar at the centre of MSH\,15-52, that was analysed in this work.
\citet{Merloni_2024} present the first data release (DR1) of eROSITA all-sky data. It includes the western Galactic Hemisphere of eRASS1, the first all-sky survey of eROSITA, including coverage of MSH\,15-52. Due to the scanning process of the all-sky survey, the exposure of the DR1 on MSH\,15-52 is lower than that of the EDR.
It does, however, grant us a wider view into the surroundings of the source. The DR1 data of MSH\,15-52 was also analysed in this work.

\subsection{H.E.S.S. public data release}

The High Energy Stereoscopic System (H.E.S.S.) telescope array is an array of five imaging atmospheric Cherenkov telescopes (IACTs) located 1800$\,$m above sea level in Namibia and sensitive to gamma rays from $\sim 0.1$ to $100 \,$TeV \citep{HESS_Crab_2006}. Initially comprised of four identical $12\,$m diameter IACTs (CT1-4), a fifth $28\,$m diameter telescope was added in 2012, improving the effective area of the array especially at lower energies \citep{HESS_CT5_2015}.
As part of the H.E.S.S. public data release \citep{hess_public_data} 9.1 hours of data of MSH\,15-52 are made public, which we used in this study.

\subsection{Fermi LAT 4FGL catalogue}

The Large Area Telescope (LAT) is the primary instrument of the Fermi satellite, sensitive to gamma rays from $20 \,$MeV to $300 \,$GeV \citep{Atwood_2009}.
In this work we made use of the fourth catalogue of Fermi LAT sources (4FGL) which contains source models, as well as flux points for a wide variety of gamma-ray sources \citep{Abdollahi_2022,fermi_4fgl}. We used flux points of MSH\,15-52 from the 4FGL catalogue as part of the SED fitting.

\subsection{Software}

In this work we used Gammapy version 1.3 \citep{Donath_2023,Gammapy_v1.3}. Furthermore the eROSITA Science Analysis Software System (eSASS), version eSASS4DR1 \citep{Brunner_2022,Merloni_2024} was used to process the eROSITA data, including data products from the calibration database (CalDB) \citep{Merloni_2024,Brunner_2022}. The Python interface of Xspec, PyXspec v. 2.1.2 \citep{Arnaud_1996}, was used as a native X-ray tool for spectral analysis.

To import X-ray models into Gammapy, a slightly modified version of the \texttt{SherpaWrapper} by \cite{Giunti_2022} was used, now maintained as part of the \texttt{gammapy-mwl}\footnote{\url{https://github.com/gammapy/gammapy-mwl}} package, to import models from the \texttt{sherpa} package v. 4.16.1 \citep{sherpa}.
For creating parallelised and reproducible data analysis workflows \texttt{snakemake} v. 9.1.3, a python-based system, was used \citep{snakemake}.

\section{eROSITA data in Gammapy}

We included both 3-dimensional (two spatial and one energy dimension) counts data and corresponding instrument response functions (IRFs) into the Gammapy framework to enable 3-dimensional fits and analysis of eROSITA data with the Gammapy software. This process was streamlined in the newly developed \texttt{eROdata} pipeline.

In this section we outline the process for all IRFs, event data, and background data. An automated pipeline was developed to conduct these conversions.
Finally a second approach is presented in which standard OGIP-format \citep{OGIP_1995} spectra can be directly read into Gammapy.

During the validation process against the eSASS command \texttt{srctool} for generating X-ray spectra \citep{Brunner_2022}, we found that there is a half-pixel shift between how eSASS and Gammapy apply the mask to the data, regardless of the size of the pixel defined by the user. The option to add this half-pixel shift was added to the \texttt{eROdata} pipeline and was used in the validation process. If exact agreement with \texttt{srctool} is not needed adding the shift is usually not necessary.
Logarithmic interpolation for both energy axes was assumed during dataset creation. This was then changed to linear interpretation before analysis.
As is done by \texttt{eSASS}, the fk5 and ICRS coordinate systems were treated as interchangeable.

\subsection{The \texttt{eROdata} framework and automated pipelines}

The \texttt{eROdata} class is a container for 3D eROSITA X-ray data that can be used to create Gammapy-compatible datasets. It comprises the following steps:

\begin{enumerate}
    \item Definition: input data, define region of interest.
    \item Event file and exposure time map creation.
    \item Exposure map creation.
    \item PSF map creation.
    \item Dataset creation.
    \item Background extraction.
    \item Stacking (optional).
    \item Rebinning (optional).
\end{enumerate}

A dataset can be created for each eROSITA TM. These datasets can also be stacked to obtain one full eROSITA dataset.
Stacking observations taken in different pointing modes was not tested and is not recommended, since averaging between considerably different exposure maps would introduce uncertainties and errors into the analysis.

All pipeline steps can be conducted manually, providing the greatest flexibility for the user.
Alternately two \texttt{snakemake} workflows are provided:
\begin{enumerate}
    \item Data workflow: creates all necessary data products but does not assemble the dataset.
    \item Full workflow: creates the full Gammapy-readable dataset.
\end{enumerate}
To use these workflows, all necessary information is provided in an input file, where certain parameters (e.g. pixel size) can be changed according to individual user needs.
The \texttt{eROdata} code can be found at the URL provided in the footnote.\footnote{\url{https://github.com/k-egg/eROdata}}

\subsection{Event lists}

\begin{figure}
   \centering
   \includegraphics[width=0.85\linewidth]{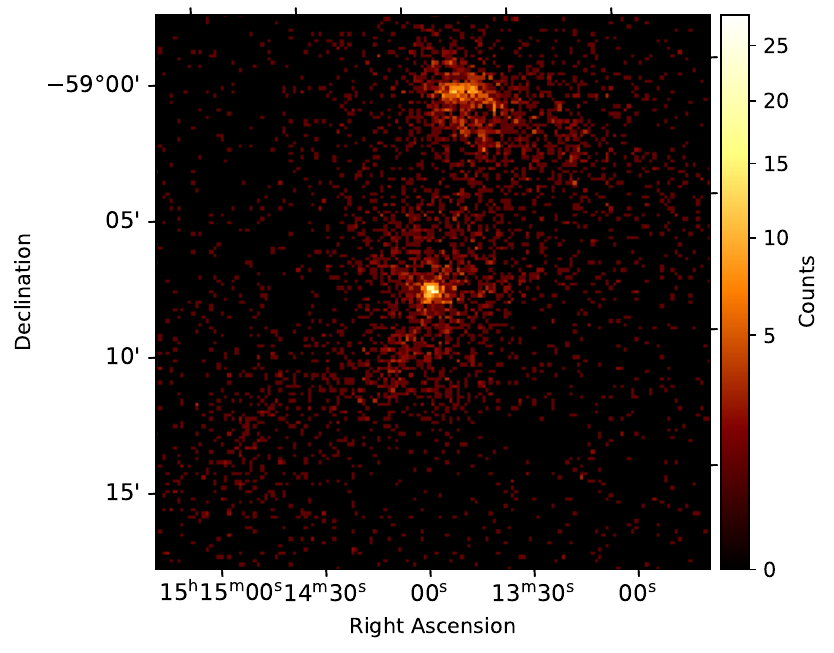}
      \caption{Map of X-ray ($\sim 0.2 - 10\, \mathrm{keV}$) eROSITA counts from MSH/,15-52 in Gammapy.}
      \label{fig:evtfile}
\end{figure}

A calibrated eROSITA event file is created by running the eSASS command \texttt{evtool}. The file contains the event list, as well as other file extensions, such as a good time interval (GTI) extension for each TM and an extension containing the pointing position of each TM over time. Optionally an image can also be generated (see Fig. \ref{fig:evtfile}).

The event list file needs to be modified for compatibility with Gammapy. A substitute `pointing direction' needs to be assigned to data not taken in pointed observations, for example DR1 survey data. For this purpose the centre of the region of interest in the analysis is chosen. This assigned substitute pointing direction should not be confused with the true pointing direction of the satellite in later analysis or diagnostic plots.
Several changes are made to the event file table and header keywords, for example the name of the reconstructed energy table column is changed from \texttt{PI}, which stands for PHA (Pulse-Height Amplitude) Invariant \citep{Arnaud_Smith_2011}, to \texttt{ENERGY}.

With these small changes Gammapy is capable of reading in the event file as an event list, including its GTI extension. These DL3 level data can then be binned into a DL4 map of counts with the regular Gammapy tools.

\subsection{Instrument response functions (IRFs)}

As photon counting experiments, the source flux of both X-ray and gamma-ray instruments cannot be directly determined from the observed number of counts, but depends heavily on the instrument response, which is characterised in IRFs. In both wavebands the forward-folding approach is used, i.e. the observed number of counts for a given model is predicted by folding it with the IRFs through \citep{Gorenstein_1967,Davis_2001,Donath_2023}
\begin{equation}
\begin{aligned}
N_\text{Pred} (p,E;\hat{\theta}) \mathrm{d}p \mathrm{d}E =& E_\mathrm{disp} \cdot [ \mathrm{PSF} \cdot (A_\mathrm{eff} \cdot t_\mathrm{obs} \cdot \Phi(p_\mathrm{true}, E_\mathrm{true}; \hat{\theta}) ] \\
    &+ \mathrm{Bkg}(p,E) \cdot t_\mathrm{obs}\,.
\end{aligned}
\label{eq:irfs}
\end{equation}
The predicted number of counts $N_\mathrm{Pred}$ over an observed position $p$ and reconstructed energy $E$ depends not only on the source flux $\Phi(p_\mathrm{true}, E_\mathrm{true}; \hat{\theta})$, exposure time $t_\mathrm{obs}$, and model parameters $\hat{\theta}$ but also on the IRFs: the energy dispersion matrix $E_\mathrm{disp}$, point spread function $\mathrm{PSF}$, effective area $A_\mathrm{eff}$, and background $\mathrm{Bkg}$.

The availability of the IRFs in a compatible format is central for data analysis, as it is required to extract meaningful, quantitative results from our data.
The following sections thus illustrate the process of converting eROSITA IRF data into Gammapy-readable formats for the construction of 3D datasets.

\begin{figure}
   \centering
   \includegraphics[width=0.45\textwidth]{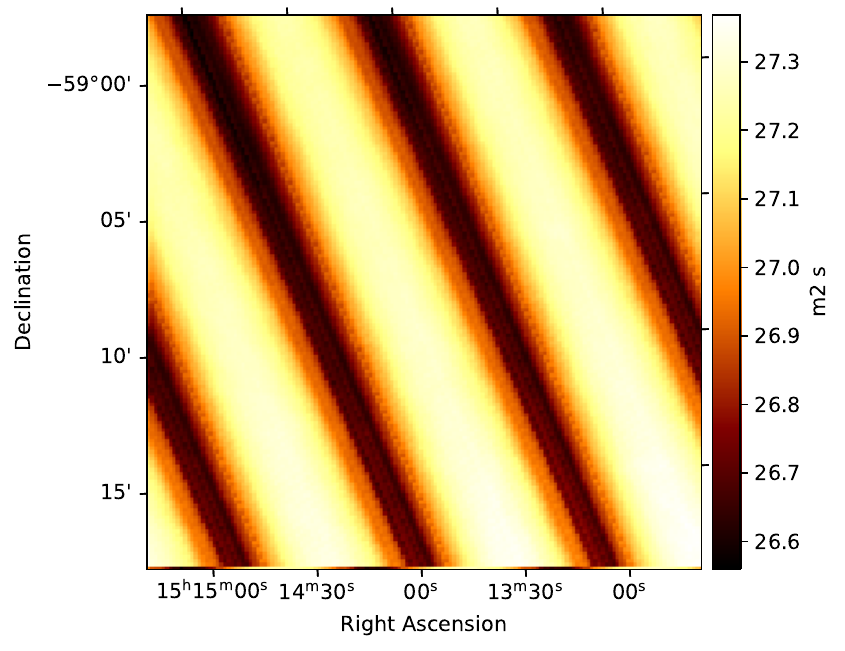}
   \includegraphics[width=0.45\textwidth]{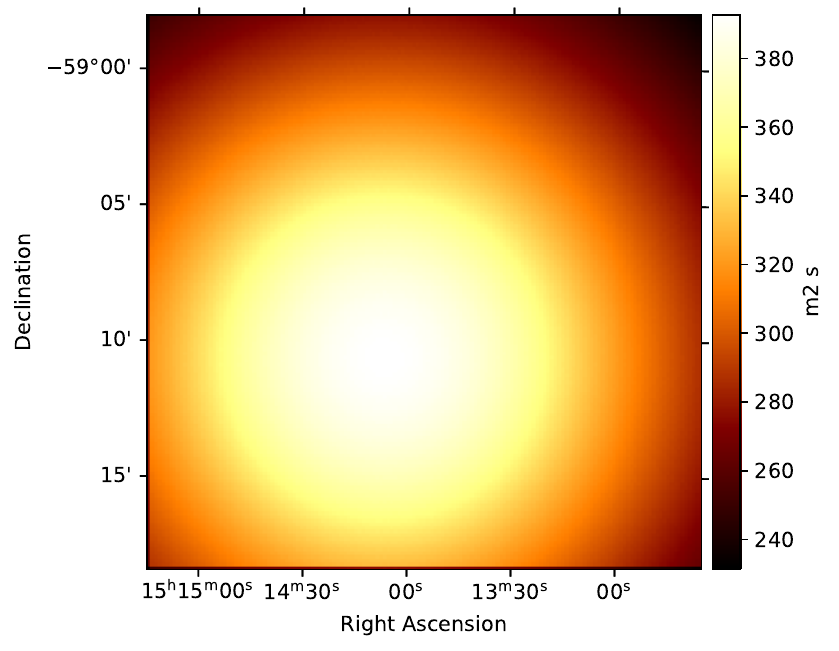}
      \caption{eROSITA example (top) DR1, (bottom) EDR exposure map at $1\,\mathrm{keV}$ shown in Gammapy.}
         \label{fig:3D_exposure}
\end{figure}

\subsubsection{Effective area}

The effective area ($A_\mathrm{eff}$) of standard eROSITA OGIP-format spectra is contained in the Auxiliary Response File (ARF) \citep{Brunner_2022}. The ARF accompanies a 1D spectrum and is valid for the specific region and time interval of the data it is extracted from.
In Gammapy datasets, the $A_\mathrm{eff}$ information is contained within the (1D or 3D) exposure map, which is $A_\mathrm{eff}$ multiplied by $t_\mathrm{obs}$ \citep{Donath_2023}.

The process of transferring the 1D $A_\mathrm{eff}$ information of the ARF into a spatially resolved 3D format was done in two steps: firstly the ARF was repeatedly extracted over small sections of, for example, $2 \times 2$ or more pixels. Lowering the resolution reduces the demands to memory and computation time. Secondly the individual ARF $A_\mathrm{eff}$ curves were assembled into a 3D $A_\mathrm{eff}$ map. Multiplied with a map of the unvignetted $t_\mathrm{obs}$ an exposure map was created. This map can be easily read into Gammapy. An example for DR1 data is shown in Fig. \ref{fig:3D_exposure}. The overlapping scans of eROSITA during its all-sky survey are clearly visible as a striped pattern, in contrast to the pointed observation (also in Fig. \ref{fig:3D_exposure}).

\begin{figure}
   \centering
   \includegraphics[width=0.45\textwidth]{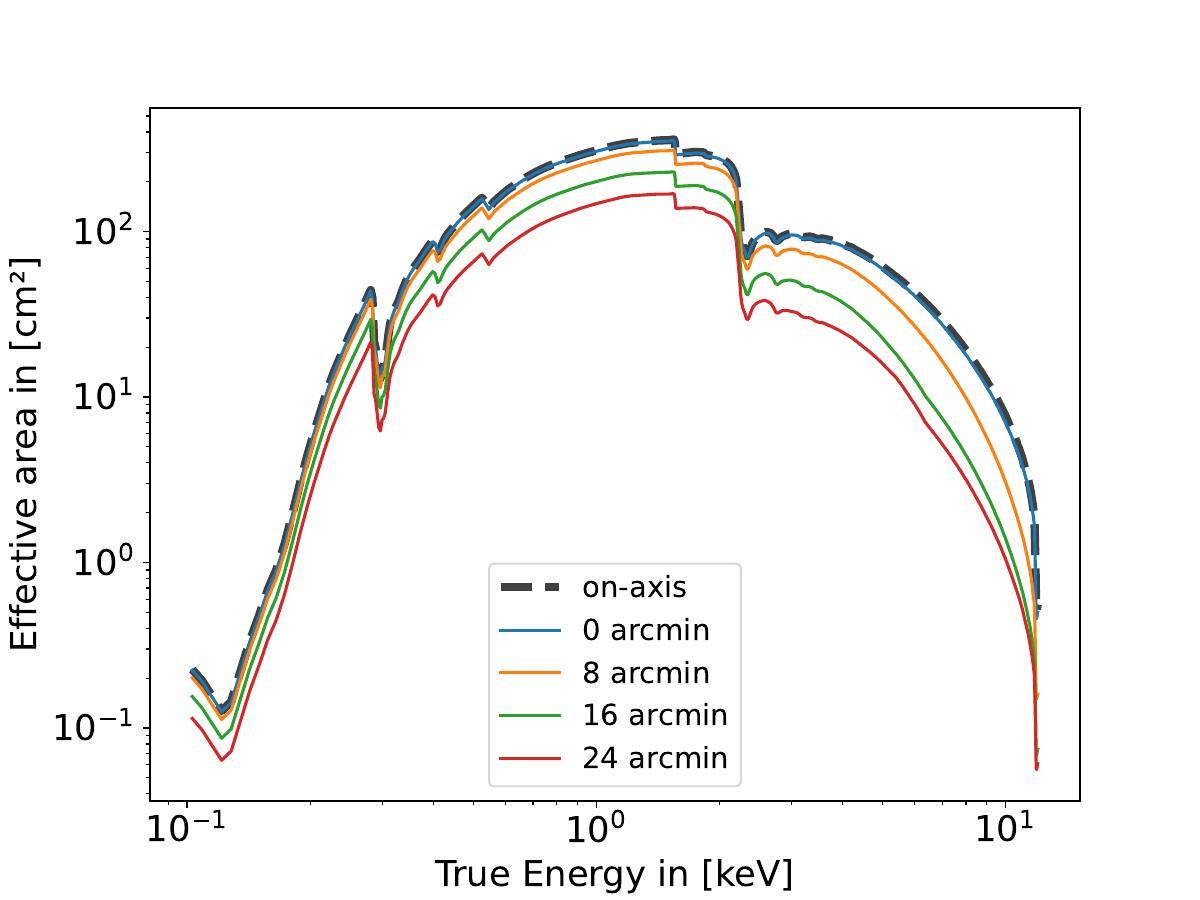}
      \caption{Pointed EDR 3D effective area map at different offsets to the centre compared to on-axis ARF.}
         \label{fig:ARF_vignetting}
\end{figure}

Fig. \ref{fig:ARF_vignetting} compares the effective area values at different points of a 3D map created for an EDR pointed observation to the CalDB on-axis ARF. The $A_\mathrm{eff}$ at the centre of the map matches the on-axis ARF, while decreasing towards the edges of the FoV.

\subsubsection{Energy dispersion}
\label{sec:IRFs:edisp}

In standard X-ray data the energy dispersion ($E_\mathrm{disp}$) information is contained within the RMF (Redistribution Matrix File). The RMF accompanies 1D spectra and contains the matrix describing the shift and crossover between the incident photon energy and the energy actually recorded by the detector \citep{Brunner_2022}. In Gammapy these energies are called `true energy' and `(reconstructed) energy' respectively \citep{Donath_2023}.

Although an RMF file is produced for each eROSITA spectrum, their content is unchanged, provided the pattern filtering of the data is consistent. The detected X-ray events are filtered for known patterns in their recorded charge clouds to ensure proper energy reconstruction \citep{Brunner_2022}. We used \texttt{pattern} 15, which includes all known shapes (single, double, quadruple). Thus only one RMF file was needed for any 3D eROSITA map.

\begin{figure}
   \centering
   \includegraphics[width=0.85\linewidth]{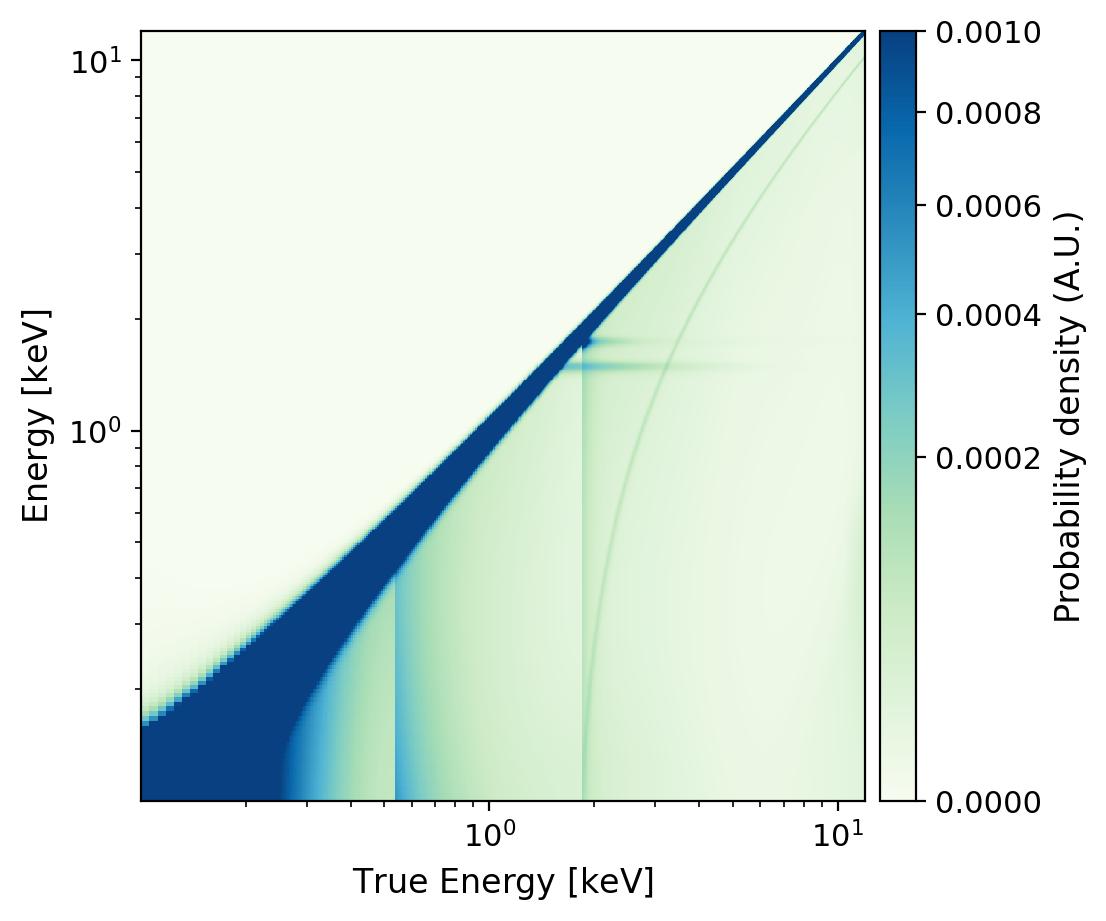}
      \caption{eROSITA RMF or EDisp matrix visualised in Gammapy.}
         \label{fig:rmf_matrix}
\end{figure}

While Gammapy possesses the functionality to read RMF files, there is a format inconsistency that necessitates a converter function.
The full-resolution eROSITA RMF matrix visualised within Gammapy is shown in Fig. \ref{fig:rmf_matrix}.

\subsubsection{Point spread function (PSF)}

While the point spread function (PSF) of eROSITA is significantly less extended than that of gamma instruments with an average half-energy width (HEW) of $\sim 26 \, \mathrm{arcsec}$, we nevertheless parametrised its influence on the spatial fit.

The eROSITA CalDB contains different versions of eROSITA 2D PSFs\footnote{\url{https://erosita.mpe.mpg.de/dr1/eSASS4DR1/eSASS4DR1_CALDB/2dpsf-doc_dr1.html}} in the OGIP 2DPSF format that were gained from calibration measurements \citep{OGIP_memo_2DPSF,Brunner_2022}. We used version 05 (\texttt{tm[1-7]\_2dpsf\_190219v05}), which is the PSF used for PSF correction in spectrum extraction.

\begin{figure}
   \centering
   \includegraphics[width=0.8\linewidth]{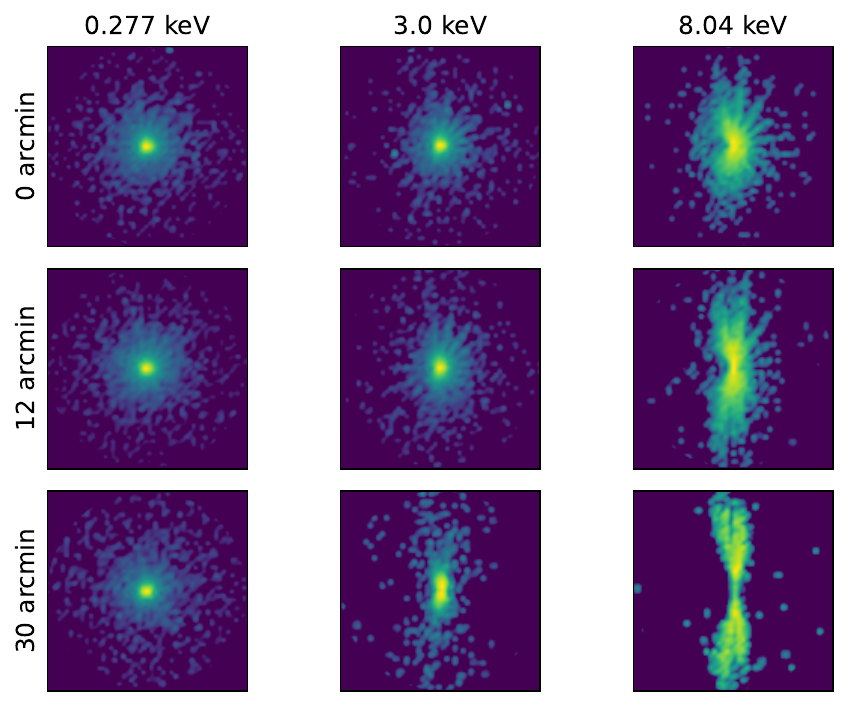}
      \caption{Example images of eROSITA's PSF for selected energies and offsets.}
         \label{fig:2D_PSF_images}
\end{figure}

Each file contains 42 images for six offset positions and seven energies over the eROSITA FoV and energy range.
A selection of these images for TM1 can be seen in Fig. \ref{fig:2D_PSF_images}.

Since Gammapy assumes radial symmetry for the PSF, the images were radially averaged and normalised. In spite of the PSF's asymmetric shape towards higher energies and towards the edge of the FoV this is a reasonable approximation, especially for survey-averaged data, which is taken over a variety of PSF positions, causing the asymmetric PSF to average out. The PSFs are provided with \texttt{eROdata} and can be read into Gammapy for use with pointed EDR data.

DR1 data from the all-sky survey, however, requires a different approach. Since Gammapy assumes a fixed pointing position, generating a map of the PSF directly for DR1 data is not possible. For this reason the PSF map creation was performed externally, analogously to the approach used for HAWC data in generating Gammapy-compatible PSF maps \citep{Olivera-Nieto_2022}.

The information on the pointing direction over time contained in the eSASS-generated event files was used to characterise the movement of the camera FoV over each pixel in the region of interest. From this an average, exposure-weighted PSF curve was calculated for each pixel and energy bin, resulting in a four-dimensional map of the PSF.

\subsection{Background}
Two main approaches to background estimation exist in X-ray analysis. Either a background spectrum can be used to profile out the background contribution from the source spectrum, an approach known as on-off background estimation in gamma-ray analysis \citep{Berge_2007}, or the background can be fit with a suitable model.
Gammapy provides the framework to realise both approaches, for gamma-ray and X-ray data.

Gammapy provides dedicated dataset frameworks for on-off data in 1D and 3D \citep{Donath_2023}. In the \texttt{eROdata} framework the background spectrum is extracted from a source-free region in the dataset. Point sources from the eROSITA DR1 one-band catalogue \citep{Merloni_2024} are excluded automatically with circular regions of $30\,$arcsec radius, since the average angular resolution of eROSITA is $26\,$arcsec. Additional exclusion regions can also be defined. The extracted background spectrum can then be used in either an on-off approach or to fit a background model.

\subsubsection{On-off approach}

For the on-off approach the background spectrum is used to fill each pixel of the map of off count, while the acceptance maps are utilised to account for the different region sizes and different exposure times of the on and off region, which is equivalent to the general X-ray procedure\footnote{\url{https://heasarc.gsfc.nasa.gov/docs/asca/abc_backscal.html}}.
Alternatively there is the option of using not only exposure time and region size, but also the effective area for weighting the background. Whilst this causes a slight increase in computation time, it can provide a more accurate background treatment for observations with larger differences in effective area (i.e. pointed observations).
Datasets created in the on-off approach should be rebinned to a minimum of five background counts per energy bin (custom function available) to ensure proper treatment of statistics \citep{Buchner2024}.

\subsubsection{Background modelling}

The background spectrum can instead be used to model the background in the region.
With eROSITA it is important to differentiate between the diffuse background (i.e. background of astrophysical origin) and the particle and instrumental background. The latter is not convolved with the exposure in a forward fit, necessitating separate treatment. The diffuse background can be described using the multi-component model by \citet{Ponti_et_al_2023}, while the particle and instrumental background is described by the filter wheel closed (FWC) model by \citet{Yeung_2023}. As the latter varies only in normalisation, a template model is provided directly with \texttt{eROdata}.

The model for the diffuse emission by \citet{Ponti_et_al_2023} can be fitted using standard X-ray tools and then imported similarly as a template model. The \texttt{eROdata} framework provides custom functions for exporting models from PyXspec into Gammapy.

\begin{figure}
    \centering
    \includegraphics[width=\linewidth]{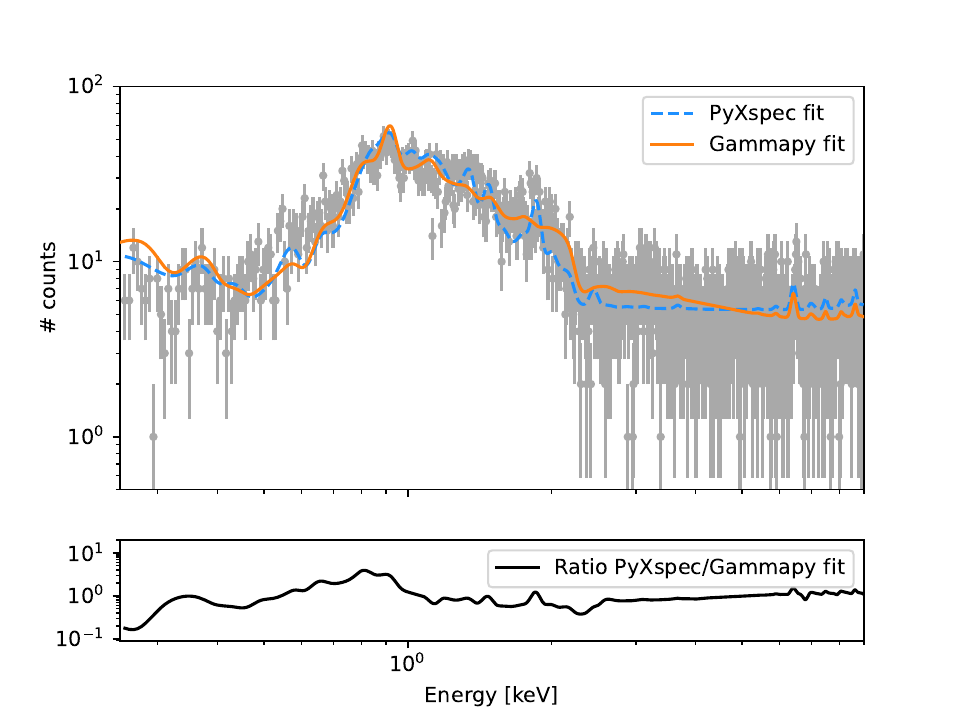}
    \caption{eROSITA DR1 background fits conducted in PyXspec and Gammapy.}
    \label{fig:background_fit_comparison}
\end{figure}

Alternatively the model can be fitted directly within Gammapy, using the \texttt{sherpa} backend. Fig. \ref{fig:background_fit_comparison} shows a background fit conducted within Gammapy compared to a background model fit in and imported from PyXspec. 
The fits found slightly different minima, likely due to different implementations of the minimisation algorithms and slight differences in the starting values, both providing a reasonable description of the background. More details on the background fit can be found in the Appendix in section \ref{sec:app:bkg}. A function that frees and adjusts necessary parameters is provided in eROdata.
Gammapy is not optimised for fitting X-ray models. Should the fit in Gammapy fail, it is recommended to import the model from PyXspec instead.

\subsection{1D OGIP spectra in Gammapy}
\label{sec:OGIP_in_Gammapy}

Gammapy natively provides a framework for reading OGIP format data. It is, however, not optimised for X-ray OGIP spectra and instead geared towards reading 1D Gamma-ray data in an internal, OGIP-similar format.
Nevertheless this tool can be utilised to read 1D X-ray spectra, an approach that is especially promising for eRASS1 catalogue spectra, which are provided in the eROSITA DR1.

To bridge the difference in format a small converter function was created, the \texttt{OGIPconverter}.
It realises all necessary format changes concerning FITS header keywords and extensions.
The RMF is converted as described in Section \ref{sec:IRFs:edisp}, whilst the ARF does not require any conversion to be readable with Gammapy.

The converter returns an energy axis reflecting the grouping of the spectrum that can be applied in Gammapy. Finally the interpolation of both energy axes can be changed to linear.

\section{1D Analysis of MSH\,15-52 in X-rays}
\label{sec:1d_analysis}

For validation of events-level eROSITA data in Gammapy, data of the MSH\,15-52 PWN region was used, both DR1 and pointed EDR data.
Counts spectra extracted with the eSASS \texttt{srctool} command were compared to 1D-spectra extracted from 3D \texttt{MapDatasetOnOff} objects in Gammapy using the custom eROdata function. A pixel size of $8\,$arcsec was chosen for the datasets. Details on the extraction regions and masks used are given in Appendix \ref{sec:app:masks}. For comparisons between counts spectra and exposure curves, see Appendix \ref{sec:app:validation}.
There is a very good agreement between the spectra in both DR1 and EDR. The ratio of exposures for the EDR data, however, drops towards higher energies. This drop can be explained by the pointed nature of the EDR observation, as well as the workings of the \texttt{srctool} command. While the overlapping scans of the DR1 data result in a more strongly averaged exposure map, the pointed exposure map encompasses a much stronger gradient over the extraction region (see Fig. \ref{fig:3D_exposure} for comparison between pointed and all-sky observations).

The \texttt{srctool} command uses a grid of points to estimate the effective area of the extraction region. Differences between this sampling method and binning of the exposure data consequently have a much higher impact on pointed observations than on all-sky observations. The impact of this on the analysis and results is discussed below.
Additionally the PyXspec analysis was compared to the analysis of eSASS-created OGIP spectra in Gammapy.

\begin{table*}
\caption{Best-fit parameters of DR1 and EDR absorbed power-law validation fits using OGIP spectra (in PyXspec), OGIP spectra (in Gammapy), 1D spectra from Gammapy 3D datasets, and 3D datasets using template models.}       
\label{tab:validation}
\centering
\begin{tabular}{l |c c c}
\hline\hline
DR1 data  &N$_\mathrm{H}$ [$10^{22}\,$cm$^{-2}$] & $\Gamma$ & $\phi_0$ [1/keV/cm$^2$/s] \\ 
\hline
   OGIP (PyXspec)  & $1.3 \pm 0.1$ & $2.0 \pm 0.2$ & $0.020 \pm 0.003$ \\
   OGIP (Gammapy)  & $1.3 \pm 0.1$ & $2.0 \pm 0.2$ & $0.021 \pm 0.003$ \\
   From Gammapy 3D & $1.3 \pm 0.1$ & $2.0 \pm 0.2$ & $0.021 \pm 0.003$ \\
   \texttt{TemplateSpatialModel} & $1.11 \pm 0.09$ & $1.72 \pm  0.12$ & $0.017 \pm 0.002$ \\
\hline\hline
EDR data &N$_\mathrm{H}$ [$10^{22}\,$cm$^{-2}$]& $\Gamma$ & $\phi_0$ [1/keV/cm$^2$/s] \\ 
\hline
   OGIP (PyXspec) &$1.01 \pm 0.02$ & $1.75 \pm 0.02$ & $0.0100\pm 0.0002$ \\
   OGIP (Gammapy) & $1.02 \pm 0.02$ & $1.76 \pm 0.02$ & $0.0101 \pm 0.0002$ \\
   From Gammapy 3D & $1.02 \pm 0.02$ & $1.76 \pm 0.02$ & $0.0101 \pm 0.0002$ \\
   \texttt{TemplateSpatialModel} & $1.09 \pm 0.01$ & $1.84 \pm  0.02$ & $0.0102 \pm 0.0002$ \\
\hline
\end{tabular}
\end{table*}

\subsection{Analysis with eSASS and PyXspec}
\label{sec:pyxspec_analysis}
Spectra generated with the eSASS \texttt{srctool} command (assuming a flat \texttt{tophat} spatial distribution and without PSF correction) were fitted using PyXspec \citep{PyXspec_2021}, which automatically subtracts the background spectra designated in the FITS headers, resulting in a fit to the background-subtracted counts spectra. An absorbed power-law was chosen as the spectral model, following \cite{Schoeck_2010}. The absorption model \texttt{TBabs} by \cite{Wilms_2000} was used with \texttt{wilm} abundances. The spectral model is

\begin{equation}
    \phi(E) = \mathrm{\texttt{TBabs}}(E,\mathrm{N}_\mathrm{H}) \cdot \phi_0 \cdot \left(  \frac{E}{E_0} \right)^{-\Gamma} \, ,
    \label{eq:abspl}
\end{equation}
where $\phi_0$ is the normalisation, $\Gamma$ the spectral index, $E_0 = 1\, \mathrm{keV}$ the reference energy, and \texttt{TBabs}$(E, \mathrm{N}_\mathrm{H})$ the energy-dependent absorption coefficient, which depends on the hydrogen column density N$_\mathrm{H}$.

The N$_\mathrm{H}$ parameter of the \texttt{TBabs} function was set to the starting value of $1.36 \cdot 10^{22}\,\mathrm{cm}^{-2}$ found in the HI4PI survey catalogue \citep{HI4PI_2016}, which was accessed through the HEASARC \texttt{nh} tool\footnote{\url{https://heasarc.gsfc.nasa.gov/cgi-bin/Tools/w3nh/w3nh.pl}}. This value also defined the maximum of the allowed parameter range. For a Galactic object, it represents the maximum absorption it can experience, whilst smaller degrees of absorption are also a possibility.

The range of data considered in the fit was set to $0.2$ to $9.0 \, \mathrm{keV}$ for the DR1 and EDR data.
The resulting best-fit parameters can be found in Table \ref{tab:validation}. 
Their errors were determined through the \texttt{error} command\footnote{\url{https://heasarc.gsfc.nasa.gov/docs/software/xspec/manual/node79.html\#error}}, which varies each parameter within its allowed limits to determine the $1 \sigma$ errors.

\subsection{Analysis with Gammapy}

\subsubsection{1D analysis of OGIP spectra in Gammapy}

Following Section \ref{sec:OGIP_in_Gammapy}, the same EDR and DR1 OGIP spectra as used in Section \ref{sec:pyxspec_analysis} were converted and read into Gammapy. The datasets were then resampled to the same grouping that is used for the PyXspec analysis.
As in Section \ref{sec:pyxspec_analysis}, they were fitted with an absorbed power-law spectral model which was imported into Gammapy using the \texttt{SherpaWrapper}. The resulting best-fit parameters are given in Table \ref{tab:validation}.

\subsubsection{1D analysis of spectra extracted from 3D Gammapy \texttt{MapDatasetOnOff} objects}

Next, for validation of the 3D eROSITA datasets, spectra were extracted from the same region (see Appendix \ref{sec:app:masks}) using the same masks that were given to eSASS.
Spectra were extracted from the stacked datasets with a custom eROdata function.
As described above, the resulting 1D spectrum was then fit with an absorbed power-law model, with best-fit parameters given in Table \ref{tab:validation}. All fit parameters agree within a 1-$\sigma$ range for DR1 and EDR.

\subsubsection{3D analysis with \texttt{TemplateSpatialModel}}
\label{sec:validation:template}

A 3D analysis was conducted on the 3D datasets using spatial template models that were extracted directly from the excess map itself. Thus the fit did not include spatial parameters and the spatial model merely functioned as a template to describe the spatial morphology of the counts. Spatial template models are a good tool for describing the complex source morphologies often found in X-rays, as shown by \citet{Picquenot_2025}. The template was multiplied with the mask used for the fit to ensure comparable results to the 1D fits. The templates for DR1 and EDR data can be seen in Fig. \ref{fig:app:validation_template} in the Appendix. As a spectral model an absorbed power-law was once again used.

\begin{figure}
    \centering
    \includegraphics[width=\linewidth]{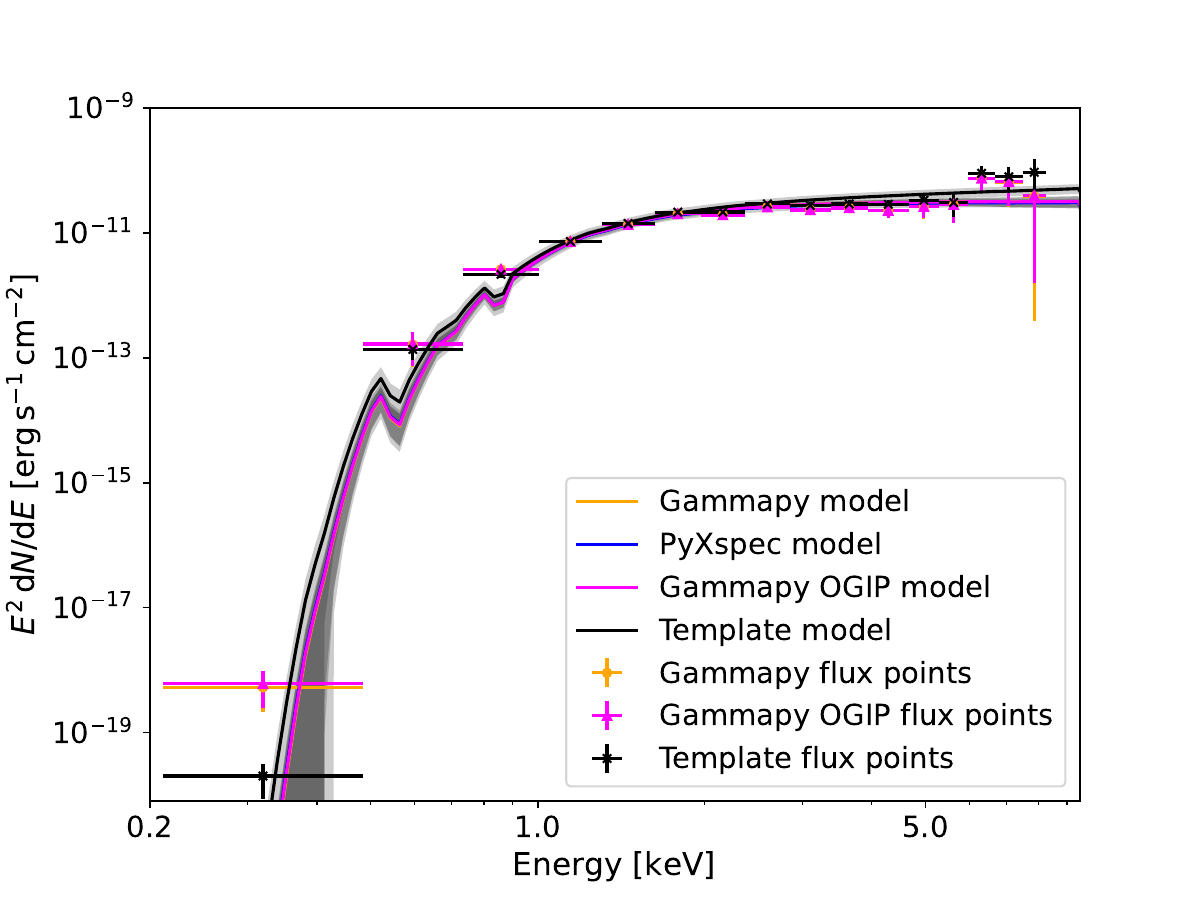}
    \caption{Comparison between absorbed power-law fits using OGIP spectra (in PyXspec), OGIP spectra (in Gammapy), 1D spectra from Gammapy 3D datasets, and 3D datasets using template models on DR1 data. All spectral best-fit models and flux points are shown, see Table \ref{tab:validation} for fit parameters.}
    \label{fig:validation_flux_plot}
\end{figure}

Fig. \ref{fig:validation_flux_plot} shows a comparison between all four fits on the DR1 data, including flux points for all fits conducted in Gammapy. The equivalent plot for the EDR data is given in Fig. \ref{fig:app:validation_flux_plot_EDR}. All 1D fits show excellent agreement with best-fit parameters within a $1 \sigma$ range. For the DR1 data the agreement with the template fit is similarly strong with best-fit parameters in the $2 \sigma$ range. For the EDR data, however, the deviations between the 1D fit and template fit are much greater, exceeding $3 \sigma$. This can be explained by the pixel-wise treatment of the exposure, which makes a far greater difference in pointed EDR data than in DR1 data. Fig. \ref{fig:validation_flux_plot} shows that the deviations between the flux points are strongest towards the lowest and highest energies considered in the fit. This might be explained by the definition of the spatial template model, which assumes a uniform morphology across all energies and the 3D exposure used in this fit. By using one template over the whole energy range a constant morphology was assumed, which can introduce systematic errors into the analysis if this does not represent the true morphology.

Deviations of $\sim\,2$ to $3\,\sigma$ can be observed between the EDR and DR1 data best-fit parameters (excluding the normalisation which is not comparable due to the different exclusion masks used). While the data possess drastically different exposure times, it is very likely that the inherent uncertainty of the 1D analysis plays a role here due to averaging over a large exposure gradient. Additionally, as only a small area towards the edge of the FoV was available for extraction of the background spectrum, contamination from the source cannot be excluded and is not further explored in this work.

\section{3D analysis of DR1 data with Gammapy}

With 3D event-level eROSITA data in Gammapy we were, for the first time, able to conduct 3D analyses in Gammapy on eROSITA X-ray data, fitting the properties of a source in both spectral and spatial dimensions simultaneously.
In this vein the DR1 eROSITA data on MSH\,15-52 were fitted with different 3D models, showcasing different possible approaches and their advantages.

\begin{table*}
\caption{Best-fit parameters of the different 3D fits on eROSITA DR1 data of MSH\,15-52.}
\label{tab:3D_DR1}      
\centering                          
\begin{tabular}{ll l c c c}       
\hline\hline                 
Source & Model & Parameter & on-off analysis & Bkg model analysis & Two-component template analysis \\
\hline                      
\textbf{MSH\,15-52}&Spectral & N$_\mathrm{H}$ [$10^{22}\,$cm$^{-2}$] & $1.4 \pm $0.1 & $1.62 \pm 0.1$ & $1.61 \pm 0.05$\\
&& $\Gamma$ & $2.04 \pm 0.14\,$ & $2.2 \pm 0.2$ &$2.18 \pm 0.09$\\
&& $\phi_0$ [1/keV/cm$^2$/s]& $0.043 \pm 0.007$ & $0.049 \pm 0.008$ &$0.041 \pm 0.003$\\
&& bkg norm. diff & - & $5400 \pm 500$ & - \\
&& bkg norm. instr. \& part. & - & $(36 \pm 4) \cdot 10^{5}$ & - \\
\hline
&Spatial & RA [deg] & $228.448 \pm 0.010$ & $228.456 \pm 0.008$ &-\\
&& DEC [deg] & $-59.108 \pm 0.008$ & $-59.114 \pm 0.006$ &-\\
&& $\sigma$ [arcmin]& $5.5 \pm 0.4$ & $5.0 \pm 0.3$ &-\\
&& $e$ & $0.89 \pm 0.02$ & $0.88 \pm 0.02$ &-\\
&& $\phi$ [deg] & $141 \pm 2$ & $142 \pm 2$ &-\\
\hline
\textbf{RCW\,89} & Spectral & norm [cm$^{-5}$] & - & - & $0.31 \pm 0.03$\\
\hline                                   
\end{tabular}
\end{table*}

\subsection{On-off analysis of DR1 data with \texttt{GaussianSpatialModel}}
\label{sec:3D:onoff}

First an on-off DR1 dataset with a pixel size of $8\,$arcsec was analysed (see Appendix \ref{sec:app:masks} for details on extraction regions and masks).
The 3D model for this fit was defined as an absorbed power-law for the spectral model, as in equation \eqref{eq:abspl}, and an elongated 2D Gaussian for the spatial model, described by\footnote{\url{https://docs.gammapy.org/2.0/user-guide/model-gallery/spatial/plot_gauss.html\#gaussian-spatial-model}}
\begin{align}
    &\phi (\mathrm{lon},\mathrm{lat}) = \frac{1}{2 \pi \sigma_\mathrm{eff}^2} \exp{\left( - \frac{1}{2} \frac{\theta^2}{\sigma_\mathrm{eff}^2}  \right)} \label{eq:2dgauss}\\
    &\text{with } \sigma_\mathrm{eff} = \sqrt{ (\sigma_M \sin{(\Delta \phi)})^2 + (\sigma_M \cdot \sqrt{1 - e^2} \cdot \cos{(\Delta \phi)})^2  }\, ,
\end{align}
where $\sigma_\mathrm{eff}$ is the effective radius of the Gaussian and $\theta$ is the separation of the evaluation point to the model centre. $\sigma_M$ defines the major axis of the Gaussian, $e$ its eccentricity, and $\Delta \phi$ the difference between the angle $\phi$ of the Gaussian and the positional angle of the evaluation point.

\begin{figure}
   \centering
   \includegraphics[width=0.8\linewidth]{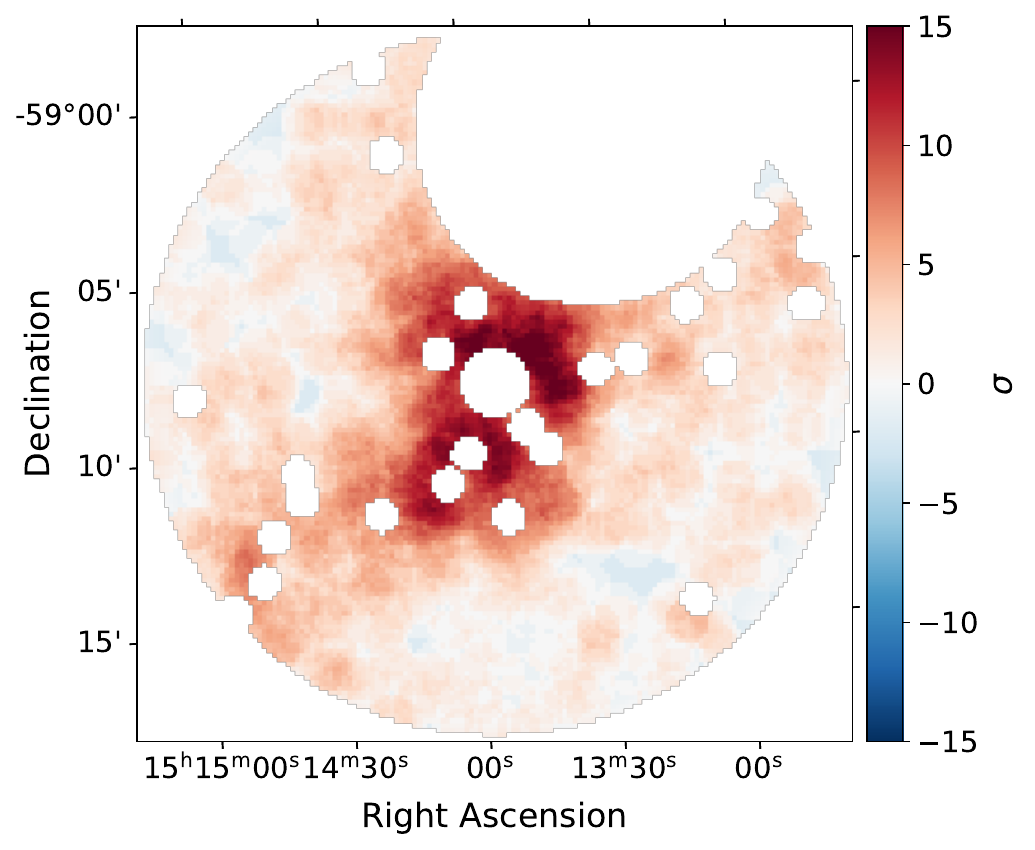}
   \includegraphics[width=0.8\linewidth]{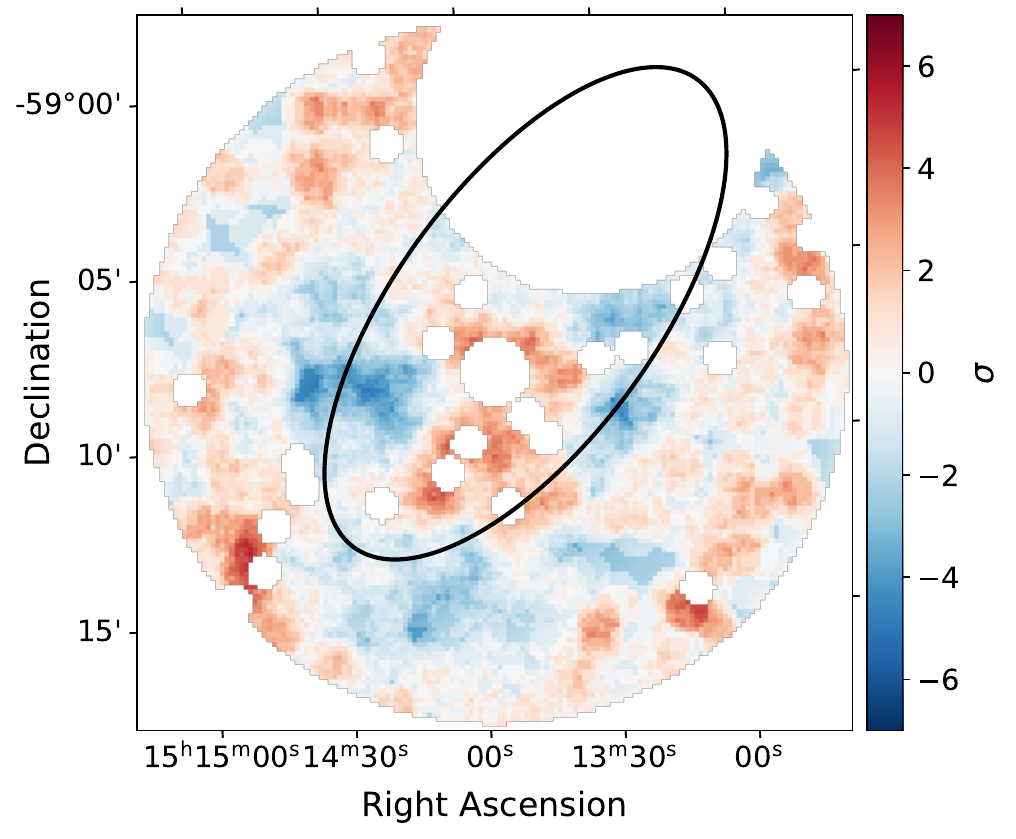}
      \caption{Significance map of MSH\,15-52 before (top) and after (bottom) the 3D fit with a 2D Gaussian spatial model, with the best-fit region overlaid on the residual map (bottom).}
      \label{fig:3D_eROSITA_sign_maps}
\end{figure}

A mask was applied to exclude all eRASS1 one-band catalogue sources (including the central pulsar) as well as RCW\,89 (see Appendix \ref{sec:app:masks}).
The fit converged, finding the best-fit parameters shown in Table \ref{tab:3D_DR1}. Fig. \ref{fig:3D_eROSITA_sign_maps} shows significance maps before and after the fit. The 1D residual significance distribution is well-described by a Gaussian with a mean of $0.1 \pm 0.03$ and 1 sigma width of $1.49 \pm 0.03$.
The 2D Gaussian spatial model provides a robust approximation for the extended emission around the central pulsar for the shallow exposure of MSH\,15-52 in the DR1 data. For deeper exposures (as in the EDR) and more complex source morphologies, the use of more detailed spatial models becomes a necessity, either by combining multiple spatial models or by using template models, which we explore below.

\subsection{Background model analysis of DR1 data}

To showcase the capabilities of Gammapy in modelling eROSITA data with a background model the 3D analysis was repeated, however this time a \texttt{MapDataset} with a pixel size of $32 \,$ arcsec was used, i.e. the background is described through models. A 1D fit of the background spectrum was conducted in Gammapy using the background model by \cite{Ponti_et_al_2023}. This model contains a description of the diffuse background and the FWC model by \citet{Yeung_2023} which describes the particle and instrumental background. The two components were treated separately in the fit. More details on the model and fit can be found in Appendix \ref{sec:app:bkg}. Both best-fit models could be saved to file and imported as template models into Gammapy with \texttt{eROdata}.

\begin{figure}
    \centering
    \includegraphics[width=\linewidth]{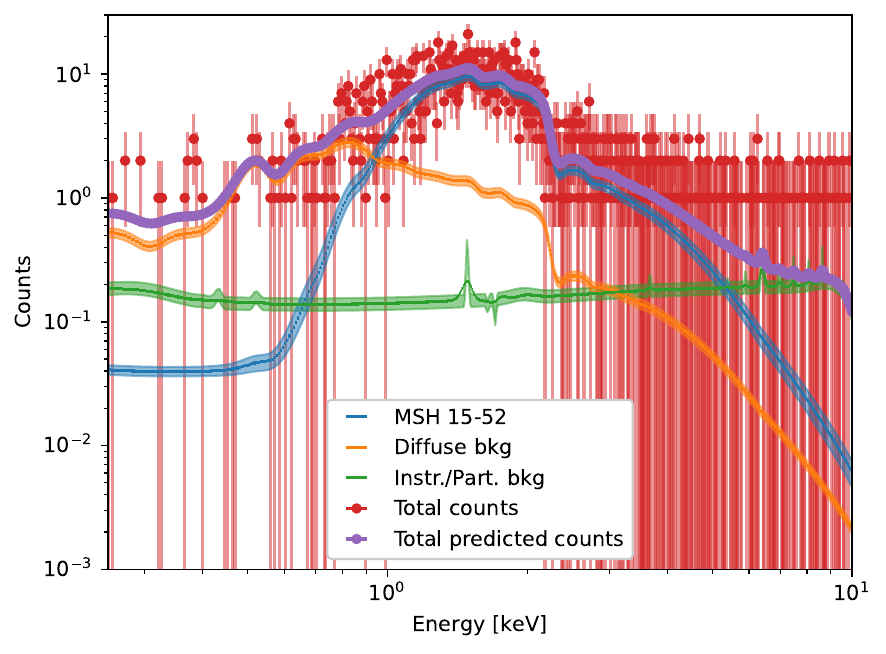}
    \caption{Spectrum of the total counts (red) and total counts predicted by the best-fit model (purple). The source model (blue), diffuse background (orange), and instrumental and particle background (green) components that make up the total predicted counts are also shown. Transparent error bands on the individual model components are shown.}
    \label{fig:background_model_npred_counts}
\end{figure}

The background model components could now be normalised on the \texttt{MapDataset} with MSH\,15-52 masked out. The region used for normalisation was identical to the background extraction region used earlier. Afterwards the normalisations of both background components were frozen and MSH\,15-52 was fit with the 3D model as described in Section \ref{sec:3D:onoff}. The results of this fit are given in Table \ref{tab:3D_DR1}, and are consistent with the on-off fit. Fig. \ref{fig:background_model_npred_counts} illustrates the fit with two background components and one source component. Error bands on the number of predicted counts for each model component were obtained by folding Gammapy's estimations in flux space with the IRFs.

\subsection{On-off two-component 3D template fit on DR1 data}
\label{sec:2comp}

Many X-ray sources possess complex morphologies that cannot be well-approximated by simple spatial models. The aforementioned spatial template models can address this issue.
For MSH\,15-52 this approach is of particular interest with respect to its overlap with the thermal source RCW\,89. While the source was previously masked out (see e.g. Fig. \ref{fig:3D_eROSITA_sign_maps}) now a simultaneous two-component modelling of both RCW\,89 and MSH\,15-52 was attempted which would present an ideal solution and allow us to disentangle the two source components. Datasets with $8\,$arcsec pixel size were used.

\begin{figure}
    \centering
    \includegraphics[width=0.49\linewidth]{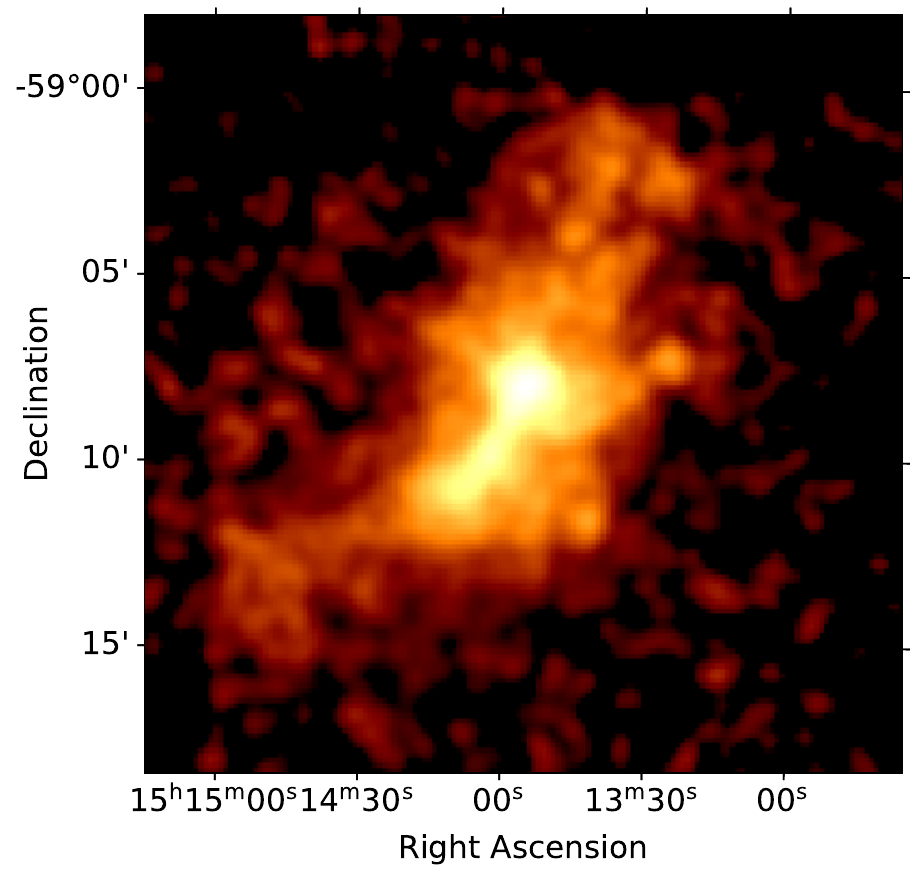}
    \includegraphics[width=0.49\linewidth]{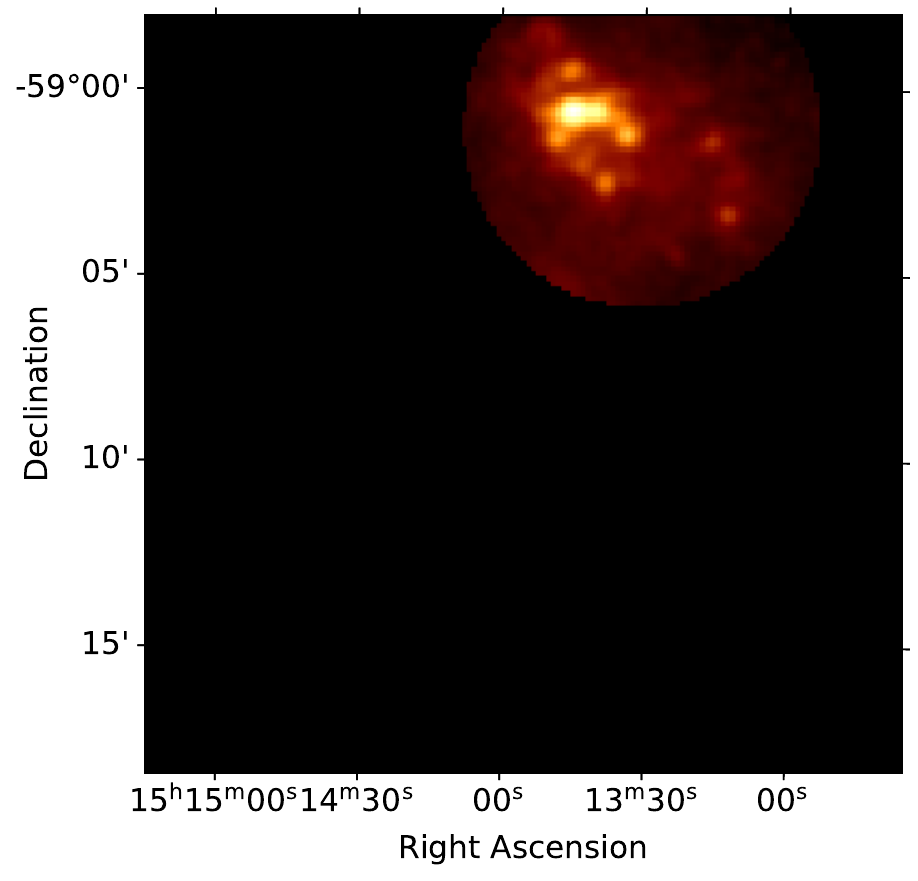}
    \caption{\texttt{TemplateSpatialModel}s for (left) MSH\,15-52 and (right) RCW\,89 generated from the EDR data.}
    \label{fig:two_comp_templates}
\end{figure}

For this purpose template models were created from the EDR MSH\,15-52 data by focusing on the energy range the respective source is brightest in, $3.0$ to $10.0 \,$keV for MSH\,15-52 and $0.2$ to $1.5\,$keV for RCW\,89. The count map slices were exposure-corrected and smoothed with a Gaussian kernel. Furthermore the 200 brightest pixels at the centre of MSH\,15-52 were replaced with average values from their vicinity. The RCW\,89 template was cut to a smaller shape by limiting its size to a circular region with a radius of $0.08\,$degrees that encompasses only RCW\,89. This was done to limit contamination from MSH\,15-52 in the template. See Fig. \ref{fig:two_comp_templates} for the template models.

Two source models were then defined: a model describing MSH\,15-52, with an absorbed power-law spectral model, and a model for RCW\,89 with an absorbed \texttt{vnei} plasma model. The RCW\,89 model description and parameters were taken from \citet{Yatsu_2005}, with only the normalisation remaining as a free parameter. Although \citet{Yatsu_2005} combined the \texttt{vnei} model with a power-law model, we assumed that the power-law component is already present through the MSH\,15-52 model. Note that the description of RCW\,89 with a single spectral model also presents an approximation, as its spectral properties vary for the different knots and structures present.

\begin{figure}
    \centering
    \includegraphics[width=0.8\linewidth]{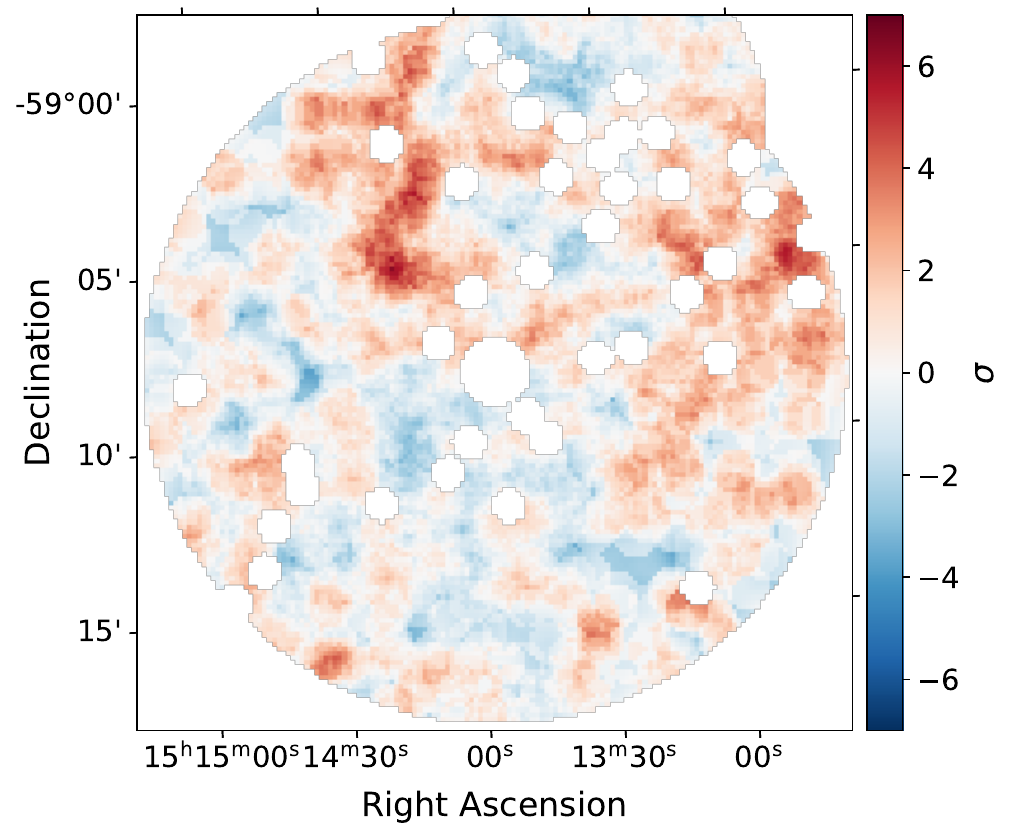}
    \caption{Significance map after the fit with a two-component \texttt{TemplateSpatialModel}.}
    \label{fig:sign_map_after_two_comp}
\end{figure}

The fit converged well, with best-fit parameters in Table \ref{tab:3D_DR1}. Fig. \ref{fig:sign_map_after_two_comp} shows the significance map after the fit, in which the emission is very well described by the model. Especially the residuals at the location of the nebula in the Gaussian fit in Fig. \ref{fig:3D_eROSITA_sign_maps} are no longer present here and the width of the residual significance distribution has decreased to $1.39 \pm 0.02$ with a centre of $0.35 \pm 0.03$.
The two emission components were effectively disentangled. This method can hence prove very powerful for other overlapping sources or crowded regions.

\subsection{Template analysis of EDR data}
\label{sec:template}

Next we take a look at the EDR data of MSH\,15-52. Due to the pointed nature of the observation, there is a significant $A_\mathrm{eff}$ gradient over the FoV which complicates the spectral analysis of this dataset with classical methods. As in Section \ref{sec:1d_analysis}, these aspects might contribute to the differences in the fit results as compared to the DR1 data. Pixel-wise treatment of the exposure and an exposure-weighted off background can help mitigate this and were hence used here.

We extracted an $8\,$arcsec-pixel size EDR dataset with an exposure-weighted off background, i.e. the background spectrum was weighted not only by the relative exposure times, but the full exposure curve. See Appendix \ref{sec:app:3d_analysis} for details.

We fitted the resulting dataset with an absorbed power-law. As a spatial model, as for the DR1 validation, a template model is used that was extracted from the excess map, smoothed with a Gaussian kernel, and multiplied with the safe mask, see Section \ref{sec:validation:template}. With this approach we collected all counts present in our data, but retained the 3D exposure and background. The best-fit parameters of this fit are in Table \ref{tab:edr_template}. When comparing the results of this fit to the 1D validation from Section \ref{sec:1d_analysis} it becomes clear that the best-fit parameters are now approaching those of the DR1 data. This speaks to an improvement in the treatment of the effective area and background, opening up new possibilities for pointed EDR eROSITA data.

\begin{table}
\caption{Best-fit parameters for EDR template fit with exposure-weighted off background.}
\label{tab:edr_template}
\centering
\begin{tabular}{l c c }
\hline\hline
Parameter & Value \\
\hline                       
N$_\mathrm{H}$ [$10^{22}\,$cm$^{-2}$]& $1.42 \pm 0.02$ \\
$\Gamma$ & $2.3 \pm 0.03$ \\
$\phi_0$ [1/keV/cm$^2$/s] & $0.0131 \pm 0.0003$ \\
\hline                                   
\end{tabular}
\end{table}
\section{Joint X-ray, GeV, and TeV analysis}

Building upon the successful 3D fits on eROSITA data we conducted a joint MWL 3D fit to eROSITA and H.E.S.S. public data \citep{hess_public_data}. Fermi LAT flux points from the 4FGL catalogue were added as a third data set \citep{Abdollahi_2022,fermi_4fgl}. Testing this application of 3D eROSITA data in Gammapy was especially crucial, as it allows us to conduct MWL 3D fits at the event level. A pixel size of $32 \,$arcsec was chosen for the eROSITA dataset.

\begin{figure}
   \centering
   \includegraphics[width=0.49\textwidth]{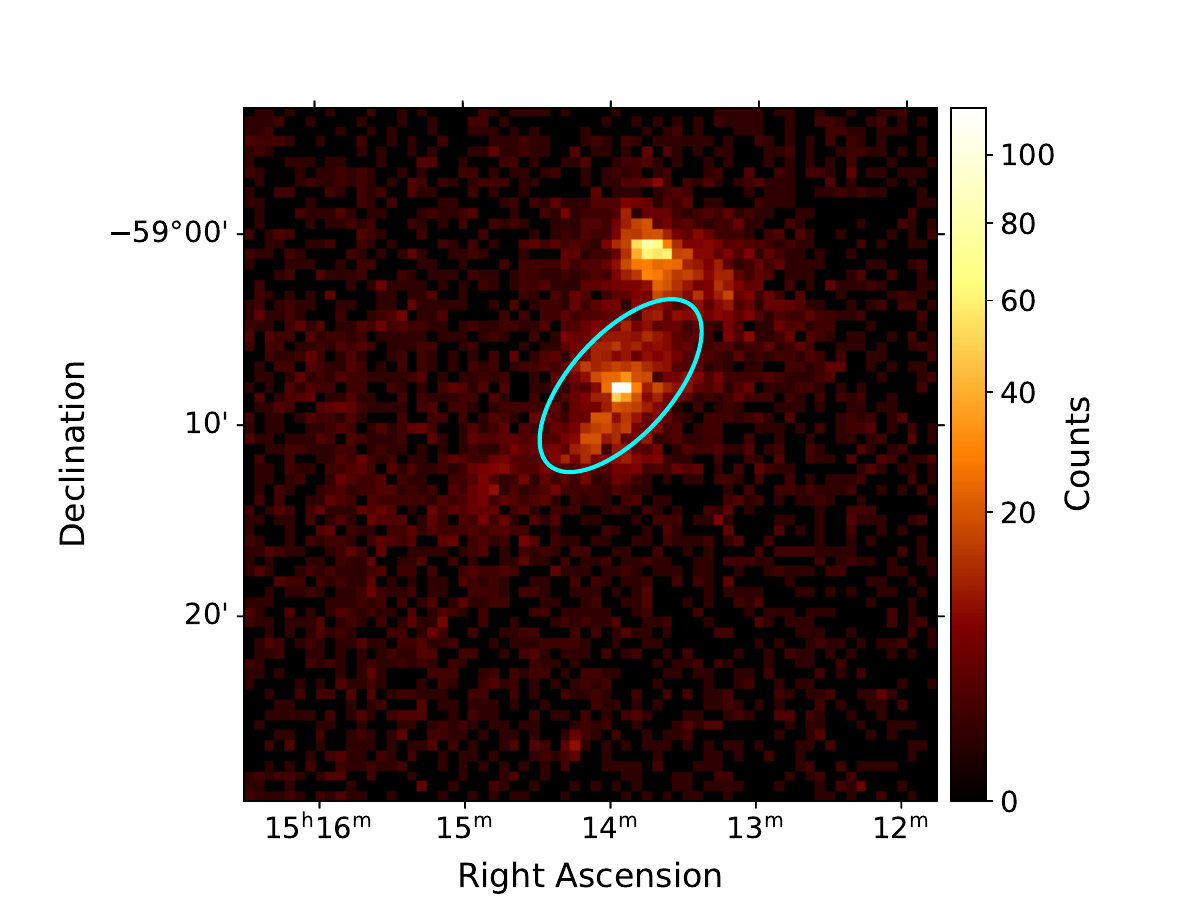}
   \includegraphics[width=0.49\textwidth]{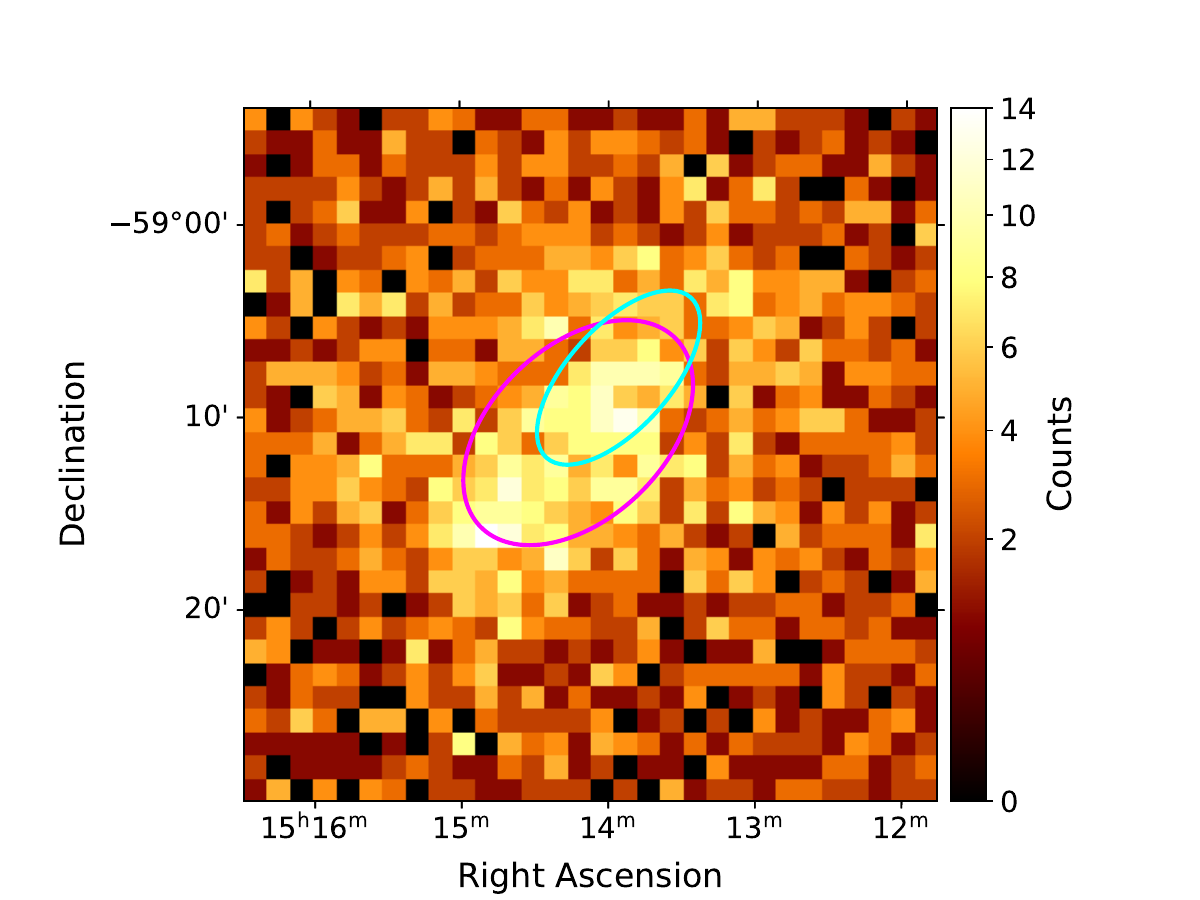}
      \caption{Counts maps of MSH\,15-52 in eROSITA (top) and H.E.S.S. (bottom) with the best-fit region of the joint fit in cyan and the best-fit region of the H.E.S.S.-only fit in magenta.}
      \label{fig:joint_fit_maps}
\end{figure}

As a spatial model, a single 2D Gaussian model was used to describe the morphology for both eROSITA and H.E.S.S. data.
To model the usual two-peak SED structure of a PWN two exponential-cutoff power-laws (ECPLs) $\phi_{1/2}$ were used. They can be expressed as\footnote{\url{https://docs.gammapy.org/2.0/user-guide/model-gallery/spectral/plot_exp_cutoff_powerlaw.html}}
\begin{equation}
\label{eq:ecpl}
    \phi_{1/2}(E) = \phi_{1/2, 0} \cdot \left(\frac{E}{E_{1/2, 0}}\right)^{-\Gamma_{1/2}} \exp(- {(\lambda_{1/2} E})^{\alpha_{1/2}}) \, .
\end{equation}
Both ECPLs were summed and photoelectric absorption was applied using the \texttt{TBabs} model as described in Section \ref{sec:pyxspec_analysis}. $\lambda_1$ was frozen, since the eROSITA data can not constrain the cut-off of the ECPL. The full spectral model is
\begin{equation}
\label{eq:joint_model}
    \phi(E) = \mathrm{\texttt{TBabs}}(E,\mathrm{N}_\mathrm{H}) \cdot \left( \phi_{1}(E) + \phi_{2}(E) \right) \, .
\end{equation}
This simplified description serves as a proof of concept for joint X-ray and gamma-ray modelling.
The fit converged with best-fit parameters given in Table \ref{tab:joint}. Fig. \ref{fig:joint_fit_maps} shows the best-fit spatial model superimposed onto the maps of eROSITA and H.E.S.S. counts. The comparison to a baseline fit on the H.E.S.S. data alone shows, that the spatial fit is strongly influenced by the eROSITA data and only provides a limited description of the H.E.S.S. data, leaving residuals after the fit. A dedicated physical analysis of MSH 15-52 will have to address different spatial morphology in X-rays and gamma-rays in more detail. Energy-dependent morphology in either energy band has, however, not been reported so far (e.g. \citet{HESS_MSH_2005,Schoeck_2010}). 

\begin{figure}
    \centering
    \includegraphics[width=\linewidth]{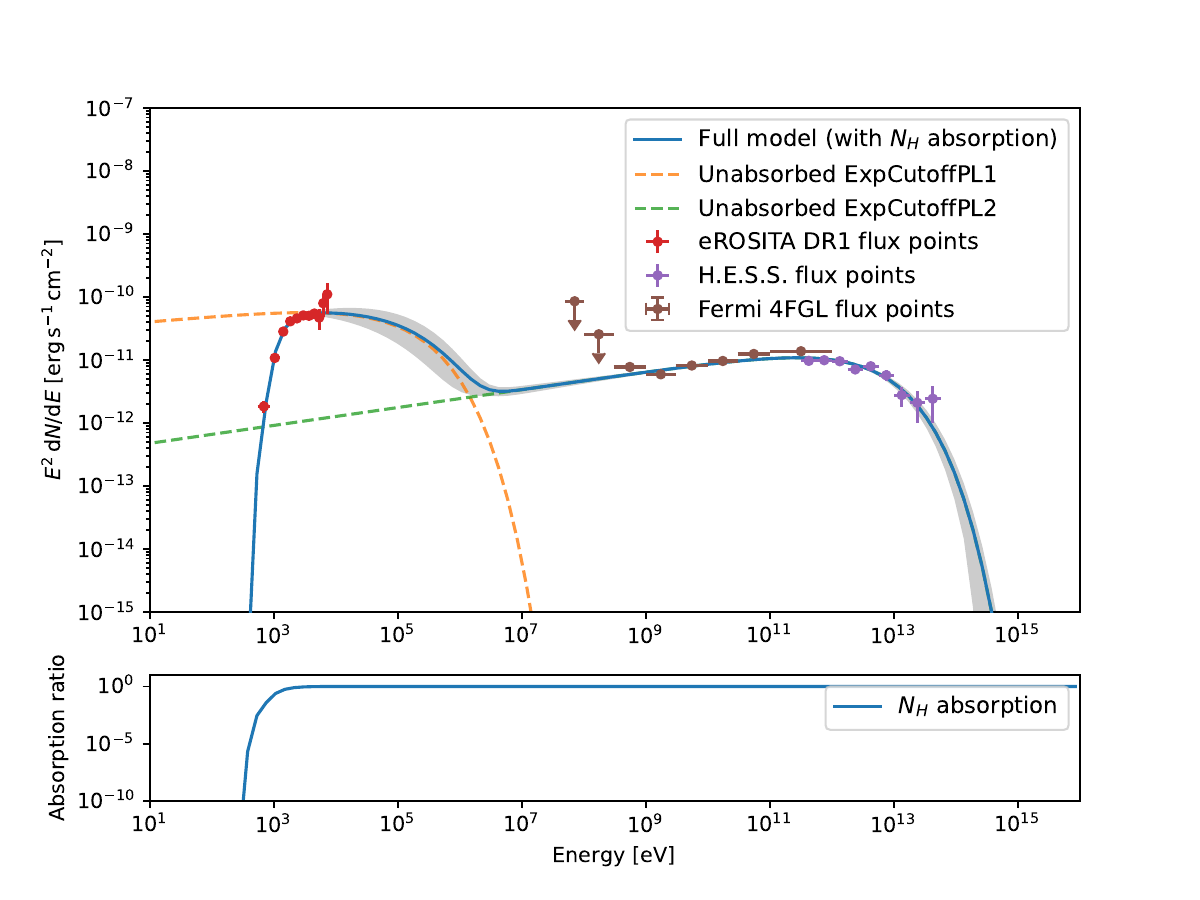}
    \caption{SED of joint MWL fit on DR1 eROSITA data, public H.E.S.S. data, and Fermi 4FGL flux points (top panel). Ratio of remaining flux after photoelectric absorption (bottom panel).}
    \label{fig:joint_sed}
\end{figure}

Fig. \ref{fig:joint_sed} shows the resulting MWL SED, with the lower panel showing the influence of photoelectric absorption on the model.
The eROSITA and H.E.S.S. flux points shown in Fig. \ref{fig:joint_sed} were derived from the best-fit model and are shown for visualisation; they were not used in the fit.

While in this non-physical description one ECPL model was mostly governed by the X-ray regime, while the other was mostly governed by the gamma-ray regime, it becomes clear that this method takes into account the interplay between them by, for instance correctly applying photoelectric absorption to the full model (see Eq. \eqref{eq:joint_model}), which ensures an accurate model prediction. A physical model description will additionally go a long way towards removing the uncertainties present in the MeV gap of the SED.
This fit thus showcases the potential of joint X-ray and gamma-ray analyses in Gammapy.

\begin{table}
\caption{Best-fit parameters joint eROSITA, H.E.S.S., and Fermi LAT fit, as defined in Eqs. \eqref{eq:ecpl}, \eqref{eq:joint_model}, and \eqref{eq:2dgauss}.}
\label{tab:joint}
\centering
\begin{tabular}{l l c }
\hline\hline
Model & Parameter & Value \\ 
\hline
Spectral & N$_\mathrm{H}$ [$10^{22}\,$cm$^{-2}$]& $1.3 \pm 0.1$ \\
& $\Gamma_1$  & $1.9 \pm 0.1$ \\
& $\phi_{0,1}$ [1/(keV s cm$^2$)]& $0.038 \pm 0.005$ \\
& $E_{0,1}$ [keV]& $1.0$ \\
& $\lambda_1$ [1/keV]& $0.01$ \\
& $\alpha_1$& $0.5$ \\
& $\Gamma_2$& $1.86 \pm 0.02$ \\
& $\phi_{0,2}$ [1/(TeV s cm$^2$)]& $1.1$e$-11 \pm 1$e$-12$ \\
& $E_{0,2}$ [TeV]& $1.0$ \\
& $\lambda_2$ [1/TeV]& $0.30 \pm 0.09$ \\
& $\alpha_2$& $0.5$ \\
\hline
Spatial & RA [deg] & $228.483 \pm 0.005$ \\
& DEC [deg]& $-59.133 \pm 0.003$ \\
& $\sigma$ [arcmin] & $5.6 \pm 0.2$ \\
& $e$ & $0.88 \pm 0.01$ \\
& $\phi$ [deg]& $138\pm 2$ \\

\hline
\end{tabular}
\end{table}

\section{Discussion}

The method and pipeline for implementing 3D eROSITA data in Gammapy that is presented in this work enables entirely new approaches to analysing MWL data. There are, however, also caveats that users should be aware of.

As seen in the validation of the EDR data, small differences in the effective area calculation between the Gammapy dataset and the eSASS data can lead to up to $5 \%$ differences between the eSASS and Gammapy 1D exposure for higher energies. This is an issue primarily for pointed EDR data, in which the FoV exposure is not averaged in the course of the scanning process of the all-sky survey.

One caveat concerning eROSITA data in general is stray light. Photons from outside the FoV can enter the camera and produce an additional background component \citep{Freyberg_et_al_2020}. This effect can be especially prominent if bright point sources are located outside of, but within $\sim 3 \, \mathrm{deg}$ of the FoV. This effect can in principle be described by an extension of the PSF \citep{Churazov_2023}. By either including such an improved PSF model in the Gammapy dataset or by subtracting it at first order, as done by \citet{Churazov_2023}, the effect of stray light can be mitigated in the Gammapy framework as well. With no bright sources in the vicinity of MSH 15-52, possible contamination was not further explored in this work.

The standard image pixel size of eROSITA is $4 \,$arcsec. Furthermore, both the reconstructed and true energy axis of eROSITA's RMF possess 1024 energy bins. As a natural consequence of this fine binning, eROSITA datasets can be very computationally expensive to create and analyse. Consequently it is recommended for users to tailor their datasets to their individual use cases and make use of helper functions that are provided in \texttt{eROdata} to reduce the file size by, for example concatenating all reconstructed energy bins outside the analysis bounds or rebinning datasets to a minimum number of background counts. Datasets can also be created with a reduced energy resolution in both reconstructed and true energy. Similarly the spatial resolution can be reduced for larger datasets, especially in the exposure map creation and for the PSF map.

The increase in bins introduced by the addition of spatial dimensions, however, brings a number of advantages. The pixel-wise treatment of the exposure and effective area is more accurate than using an average, especially in regions with high gradients in the exposure, such as pointed observations and does not require any foreknowledge of the shape of the emission. The same effect can be seen in the calculation of on-off backgrounds. Weighting the background counts not only by exposure time, but also by effective area can provide a much more accurate background treatment in on-off analysis.
Background fitting is also possible within Gammapy. By defining spatially constant models for the diffuse background and particle and instrumental background, the background can be separated from the source component, which is assigned a compact model or template.
Furthermore the 3D approach also enables the separation of separate sources and source components through the simultaneous fit of several, even overlapping, three-dimensional model components.

The biggest advantage, however, remains the ability to conduct 3D MWL fits at photon event level. Gammapy's flexible framework ensures that all manner of datasets can be included, 3D, 1D, or even datasets consisting of flux points, as was shown in the MWL analysis conducted in this work. Including as much event level data as possible leads to better treatment of systematics, background, and photoelectric absorption. Especially when combined with physical models from the \texttt{naima} \citep{naima_2015} and \texttt{Gamera} \citep{Hahn_2016} packages, as done in \citet{eROSITA_ICRC_2025}, as well as the nested sampling algorithms that have been included into Gammapy since version 2.0. this unlocks a powerful new approach to MWL analysis.
With the work of \citet{Rosillo_2025} on importing 1D spectra over many orders of magnitude into Gammapy and the work of \citet{Unbehaun_2024} on including KM3NeT neutrino data into Gammapy, Gammapy has the potential to become a highly adaptable tool for MWL and multi-messenger studies.

\section{Conclusions}

In this work we conducted a joint analysis of three-dimensional X-ray and gamma-ray data. We present the \texttt{eROdata} pipeline for converting eROSITA data into a Gammapy-readable three-dimensional dataset format. Both event and IRF data were converted and concatenated into a dataset.
The approach was validated through comparison with standard eROSITA data products in the spectral fitting tool PyXspec, finding excellent agreement within $1 \,\sigma$ for 1D fits in PyXspec and Gammapy. The agreement with 3D fits using template models is within $1\, \sigma$ for DR1 data and within 2 to $3 \, \sigma$ for EDR data.

We show that 3D fits on X-ray data are possible, and even complex source morphologies can be described and disentangled through the use of template models. Different 3D fitting approaches are presented using a modelled background, an exposure-weighted on-off background, and different spatial models. The fits agree within $1$ to $2 \, \sigma$.

Finally we conducted a joint 3D MWL fit using 3D eROSITA data, 3D H.E.S.S. public data, and 4FGL catalogue Fermi LAT flux points. This demonstrates that 3D MWL fits over a broad energy range are possible and can be used to characterise MWL SEDs. This will help us to explore the properties of not only PWNe but also many other objects bright over many orders of magnitude in energy.

\section*{Data availability}

The \texttt{eROdata} scripts and workflows for converting and assembling eROSITA 3D datasets can be found under \url{https://github.com/k-egg/eROdata} and at \url{https://doi.org/10.5281/zenodo.21511138}.

\begin{acknowledgements}
We thank the members of the Gammapy developers team and the H.E.S.S. collaboration, alongside colleagues at ECAP, for their valuable feedback and discussions.\\

K. Egg and A. M. W. Mitchell are supported by the Deutsche Forschungsgemeinschaft, DFG project number 452934793.\\

This work made use of Gammapy \citep{Donath_2023}, a community-developed Python package. The Gammapy team acknowledges all Gammapy past and current contributors, as well as all contributors of the main Gammapy dependency libraries: \href{https://numpy.org/}{NumPy}, \href{https://scipy.org}{SciPy}, \href{http://www.astropy.org}{Astropy}, \href{https://astropy-regions.readthedocs.io}{Astropy Regions}, \href{https://scikit-hep.org/iminuit}{iminuit}, \href{https://matplotlib.org}{Matplotlib}.\\

      This work is based on data from eROSITA, the soft X-ray instrument aboard SRG, a joint Russian-German science
mission supported by the Russian Space Agency (Roskosmos), in the interests of the Russian Academy of Sciences
represented by its Space Research Institute (IKI), and the Deutsches Zentrum für Luft- und Raumfahrt (DLR). The
SRG spacecraft was built by Lavochkin Association (NPOL) and its subcontractors, and is operated by NPOL with
support from the Max Planck Institute for Extraterrestrial Physics (MPE). The development and construction of the
eROSITA X-ray instrument was led by MPE, with contributions from the Dr. Karl Remeis Observatory Bamberg \&
ECAP (FAU Erlangen-Nuernberg), the University of Hamburg Observatory, the Leibniz Institute for Astrophysics
Potsdam (AIP), and the Institute for Astronomy and Astrophysics of the University of Tübingen, with the support of
DLR and the Max Planck Society. The Argelander Institute for Astronomy of the University of Bonn and the Ludwig
Maximilians Universität Munich also participated in the science preparation for eROSITA.\\
The eROSITA data shown here were processed using the eSASS software system developed by the German
eROSITA consortium.\\

This work made use of data from the H.E.S.S. DL3 public test data release 1 (HESS DL3 DR1, H.E.S.S. collaboration, 2018).
\end{acknowledgements}

\bibliographystyle{aa}
\bibliography{bibliography.bib}

\begin{appendix}
\onecolumn
\clearpage

\section{Extraction regions for MSH\,15-52 spectra and point source exclusion}
\label{sec:app:masks}

Most regions of interest in eROSITA data contain unrelated point sources. Defining additional source models for all of them was, however, not possible. The soft X-ray energy regime of eROSITA encompasses thermal effects and spectra that can usually not all be modelled in the scope of an analysis. In order to limit contamination, we thus defined masks for point source exclusion.

We used the eRASS1 1-band catalogue sources \citep{Merloni_2024} to define exclusion regions. For the pointed EDR data of MSH\,15-52 we excluded additional point sources by eye, as no catalogue for these data was made public.
Additionally RCW\,89 was excluded in most DR1 analyses with a circular region with a radius of $0.08\,$degrees, while $1.2\,$degrees were used for the EDR data. See Fig. \ref{fig:app:validation_template} for mask contours.
To characterise the background, an annulus region around the source region was chosen, $0.5\,$degrees (until the edge of the pointed FoV) for the EDR data and $0.7\,$degrees for the DR1 data.

\section{Validation fits}
\label{sec:app:validation}

\begin{figure*}[h]
   \centering
   \includegraphics[width=0.49\columnwidth]{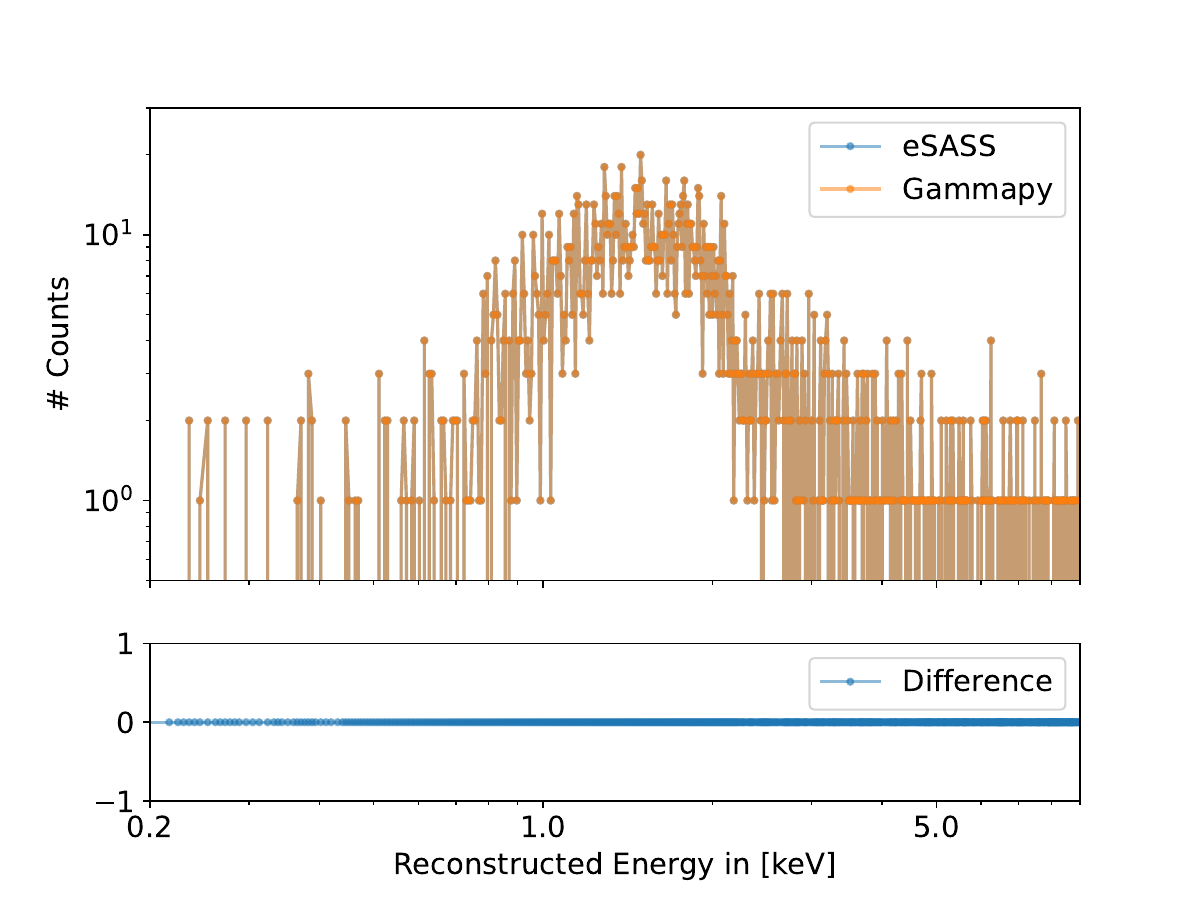}
   \includegraphics[width=0.49\columnwidth]{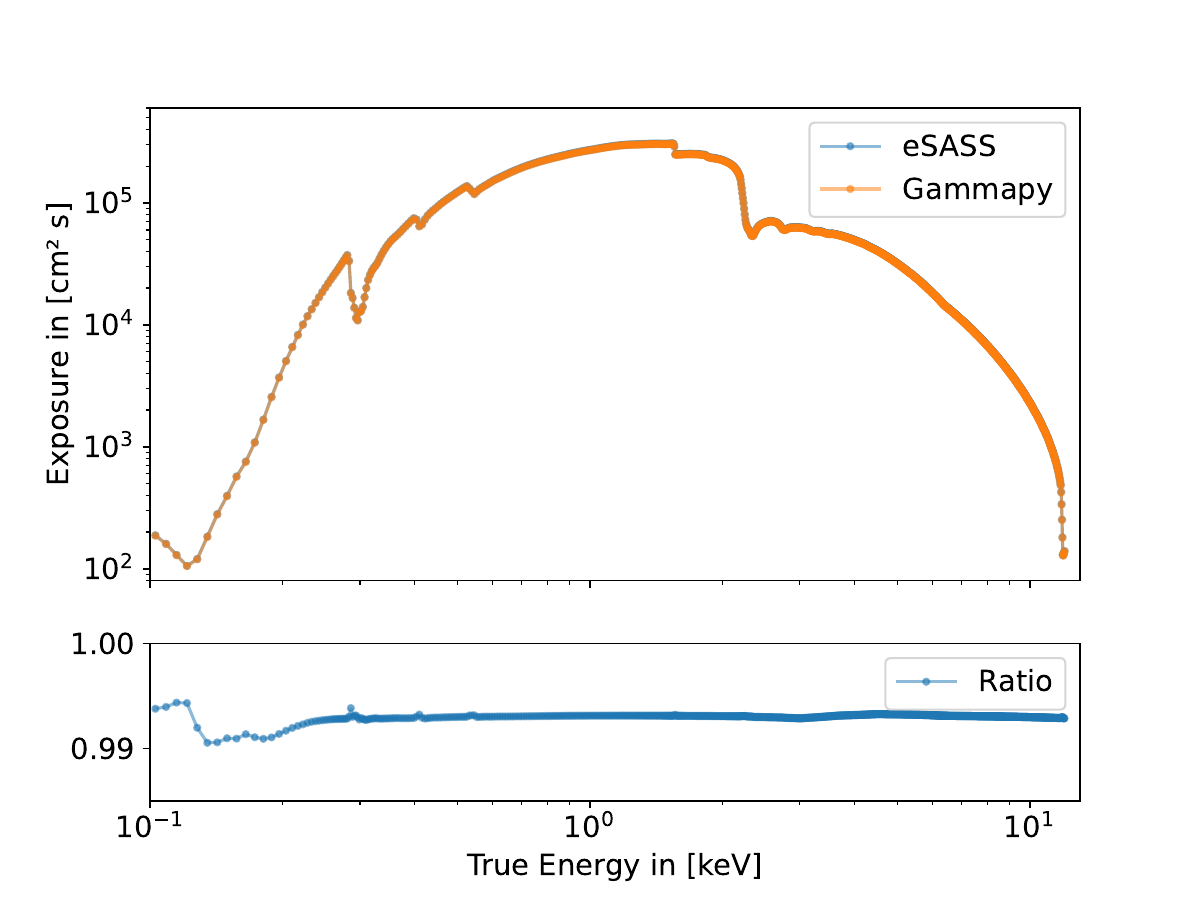}
      \caption{Comparison between a DR1 stacked or 820 1D spectrum (left) and exposure curve (right) created with eSASS and inside Gammapy.}
      \label{fig:app:spec_comp_DR1}
\end{figure*}

\begin{figure*}[h]
   \centering
   \includegraphics[width=0.49\columnwidth]{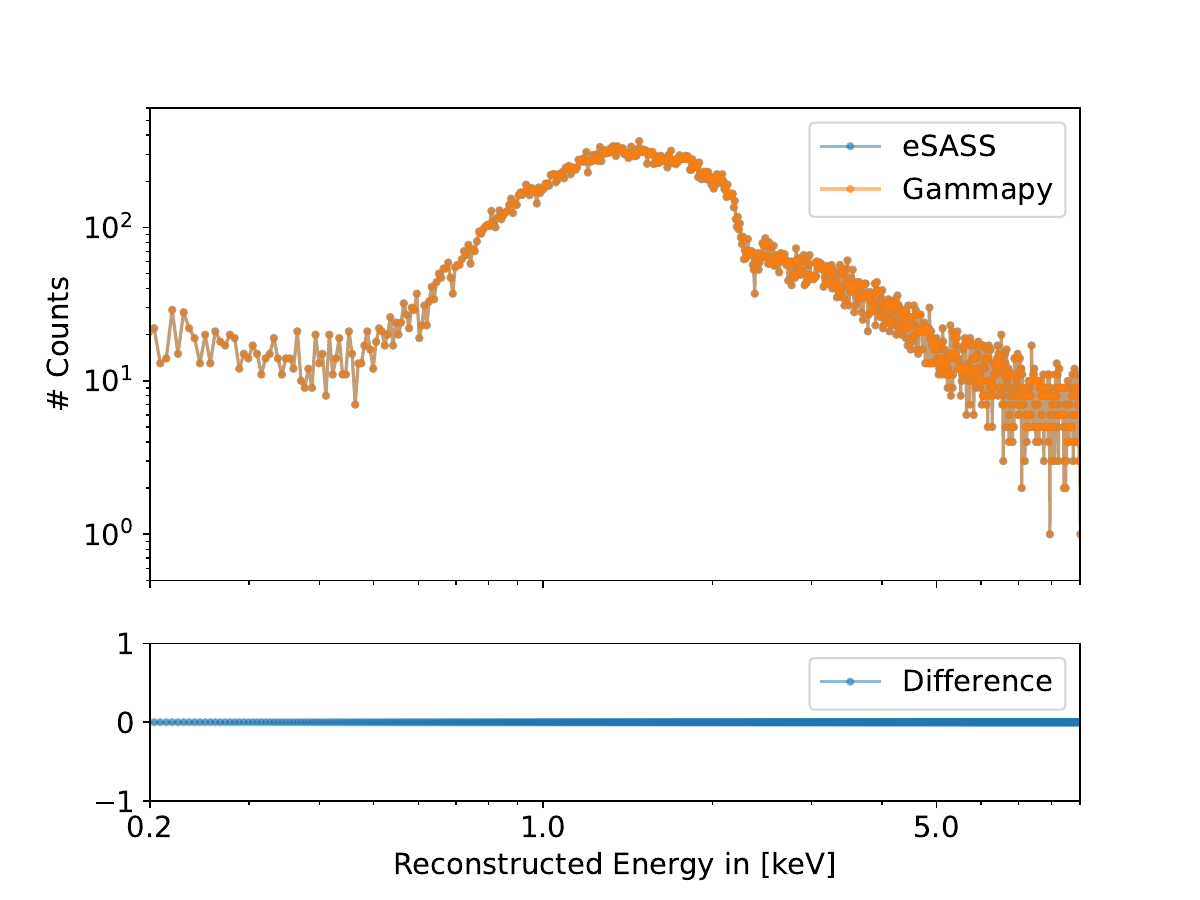}
   \includegraphics[width=0.49\columnwidth]{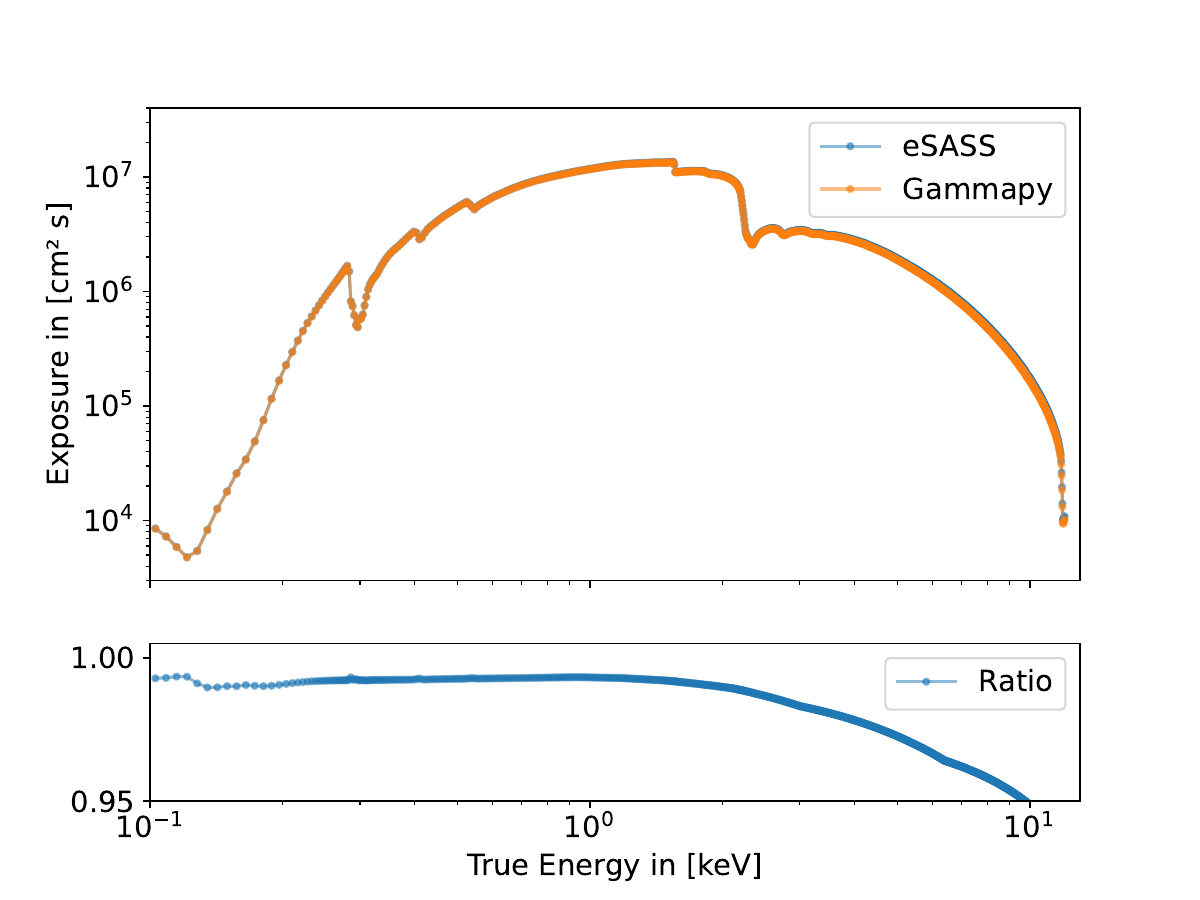}
      \caption{Comparison between an EDR stacked or 820 1D spectrum (left) and exposure curve (right) created with eSASS and inside Gammapy.}
      \label{fig:app:spec_comp_EDR}
\end{figure*}

\begin{figure*}[h]
    \centering
    \includegraphics[width=0.49\columnwidth]{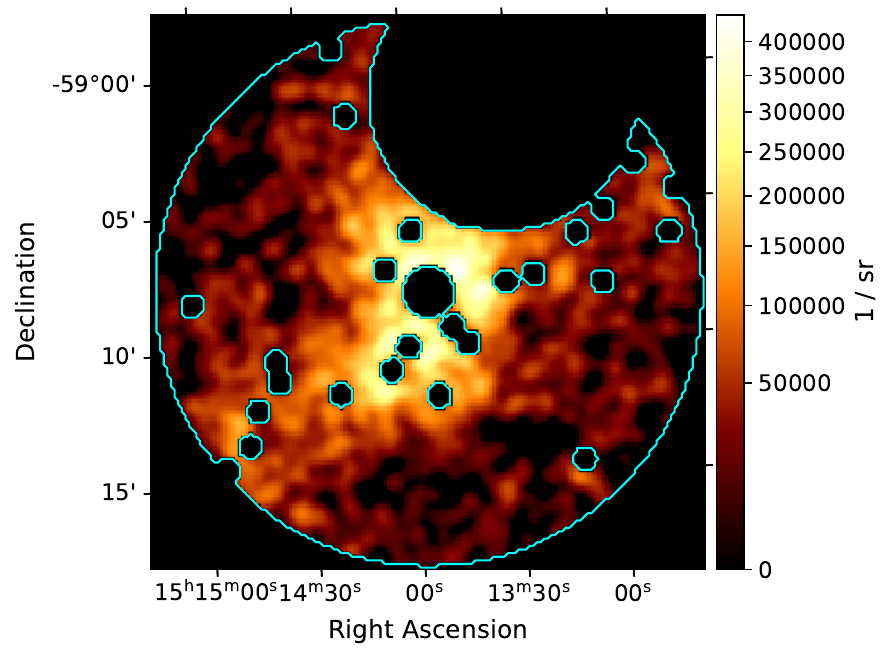}
    \includegraphics[width=0.49\columnwidth]{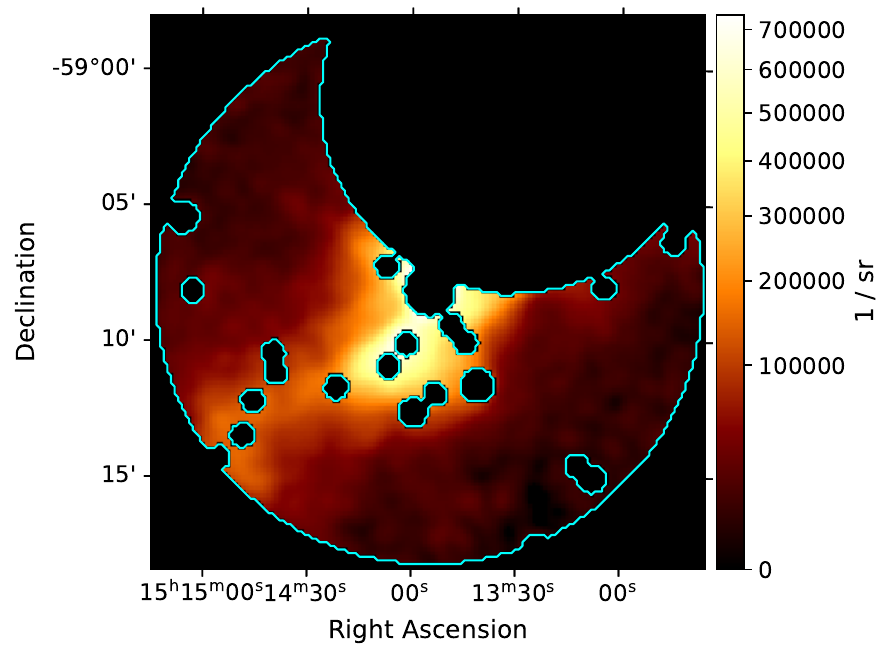}
    \caption{\texttt{TemplateSpatialModel} derived from (left) DR1 and (right) EDR data of MSH\,15-52, mask contours are outlined in cyan.}
    \label{fig:app:validation_template}
\end{figure*}

\begin{figure}[!h]
    \centering
    \includegraphics[width=0.49\linewidth]{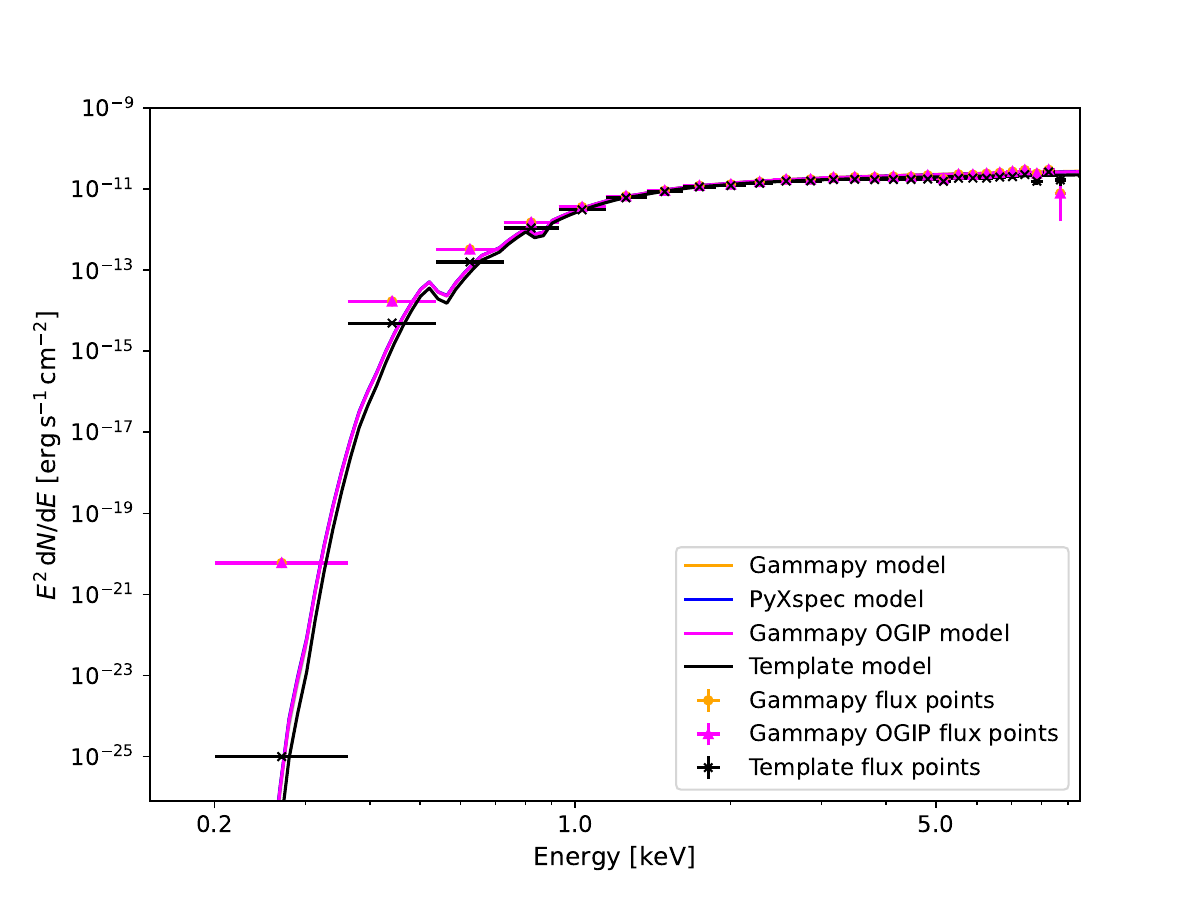}
    \caption{Comparison between eSASS, eSASS (in Gammapy), Gammapy 1D, and \texttt{TemplateSpatialModel} absorbed power-law fits on EDR data. All spectral best-fit models and flux points are shown, see Table \ref{tab:validation} for fit parameters.}
    \label{fig:app:validation_flux_plot_EDR}
\end{figure}

The direct comparison between 1D counts spectra and exposure curves created with eSASS and within Gammapy can be seen in Fig. \ref{fig:app:spec_comp_DR1} for DR1 data and in Fig. \ref{fig:app:spec_comp_EDR} for EDR data. The eSASS exposure curve was obtained by multiplying the effective area curve with the exposure time.
The \texttt{TemplateSpatialModel}s used in the template validation fits can be seen in Fig. \ref{fig:app:validation_template}.
A plot comparing the validation models fitted to the EDR data and the flux points derived from them can be found in Fig. \ref{fig:app:validation_flux_plot_EDR}.

\FloatBarrier

\section{Background fit}
\label{sec:app:bkg}

The eROSITA background can be split into two main components: the diffuse X-ray background and the instrumental and particle background \citep{Merloni_2024}.
The diffuse X-ray background can vary on a case-by-case basis. It is made up of X-ray photons that are emitted by diffuse background components present in the X-ray sky. It is characterised and modelled by \cite{Ponti_et_al_2023} who find six components (not including Solar wind charge-exchange (SWCX) in this fit). The local hot bubble (LHB) contributes on local scales, the circumgalactic medium (CGM) and hot corona (Cor) on Galactic scales, while the cosmic X-ray background (CXB) describes the unresolved background of active galactic nuclei (AGNs) and galaxies. Each component was modelled with a model component: the LHB, CGM, and Cor are described with \texttt{apec} models, models for collisionally ionised diffuse gas\footnote{\url{http://www.atomdb.org/}} \citep{2001_apec}; the CXB was modelled with a power-law. Additionally photoelectric absorption via the \texttt{TBabs} model \citep{Wilms_2000} was applied to the CGM, Cor, and CXB. A constant factor was added for ease of fitting, containing the size of the fitted region in arcmin$^2$. For Gammapy the readily available size of the unmasked extraction region of the spectrum was used instead. The \texttt{apec} model is created from atomic table data and was adjusted in the fit using the free parameters given in Table \ref{tab:app:bkg_fit}. Since the \texttt{acx2} model could not be imported into Gammapy, it is not used in this work.
The total model for the diffuse background thus amounts to
\begin{equation}
\label{eq:app:erosita_backgr_gp}
    \mathrm{\texttt{const}*} \mathrm{(}\mathrm{\texttt{apec} + \texttt{TBabs}*(\texttt{apec} + \texttt{apec} + \texttt{powerlaw}))}\, .
\end{equation}

The particle and instrumental background of eROSITA, meanwhile is made up of, for example, charged particles and single-reflected X-rays \citep{Freyberg_et_al_2020}. It is modelled by \cite{Yeung_2023}. Only the normalisation of this model is expected to change and the model is not folded with the effective area during the fit, owing to the fact that the particles hit the detector without adhering to the normal photon path and detector geometry.

The two main model components were fit in PyXspec, as well as in Gammapy. The resulting best-fit parameters can be found in Table \ref{tab:app:bkg_fit}. A plot of both fits can be seen in Fig. \ref{fig:background_fit_comparison}.
Model parameters were constrained following \citet{Ponti_et_al_2023}, while the kT parameter of the LHB was estimated from \citet{Liu_et_al_2016}. For the fit within Gammapy the \texttt{sherpa} backend was used. This fitting backend does not provide uncertainties on the best-fit values. While the fitted model can thus provide a robust description of the present background emission, no further conclusions on the background components and emission should be drawn from the fit at present. Table \ref{tab:app:bkg_fit} shows that the fitted parameters do not fully match between PyXspec and Gammapy. These differences may arise from the rather complicated model and different implementations of fitting algorithms used. However, with no estimated uncertainties on the Gammapy best-fit it is difficult to quantify these differences. Users are advised to check the background fit with standard X-ray fitting tools if necessary.

\FloatBarrier

\begin{table*}[h!]
\caption{Best-fit values of the eROSITA background fit in PyXspec and Gammapy.} 
\label{tab:app:bkg_fit} 
\centering

\begin{tabular}{l|c | c | c }       
\hline\hline
Model component & Parameter name & Best-fit value PyXspec & Best-fit value Gammapy \\[0.1cm]
\hline
 \texttt{const} (Area normalisation)& factor [arcmin$^2$] & 5170.40  & 5541.76\\[0.1cm]
 \hline
 \texttt{apec} (LHB)& kT [keV] & 0.1 & 0.1\\[0.1cm]
 & Abundanc & 1.0 & 1.0\\[0.1cm]
 & Redshift & 0.0 & 0.0\\[0.1cm]
 & norm [cm$^{-5}$] & $1.2 \mathrm{e-06}^{+1\mathrm{e}-07} _{-1\mathrm{e}-07} $ & $2.76\mathrm{e}-06$ \\[0.1cm]
 \hline
 \texttt{TBabs} (N$_\mathrm{H}$ absorption)& nH [$10^{22}$cm$^{-2}$]& 1.36 & 1.36 \\[0.1cm]
 \hline
\texttt{apec} (CGM) & kT [keV]& $0.150^{+0.001} _{-0.0} $ & 0.15\\[0.1cm]
 & Abundanc & $0.08^{+0.02} _{-0.02} $ & 0.040 \\[0.1cm]
 & Redshift & 0.0 & 0.0\\[0.1cm]
 & norm [cm$^{-5}$]& $0.008^{+0.001} _{-0.001} $ & 0.0132 \\[0.1cm]
\hline
\texttt{apec} (Galactic Corona) & kT [keV]& 0.7 & 0.7\\[0.1cm]
 & Abundanc & 1.0 & 1.0\\[0.1cm]
 & Redshift & 0.0 & 0.0\\[0.1cm]
 & norm [cm$^{-5}$]& $8\mathrm{e}-06^{+2\mathrm{e}-06} _{-2\mathrm{e}-06} $ & $1.00\mathrm{e}-08$ \\[0.1cm]
 \hline
\texttt{powerlaw} (CXB) & PhoInd & 1.46 & 1.46 \\[0.1cm]
 & norm [1/keV/cm$^2$/s]& $3.6 \mathrm{e}-06^{+3\mathrm{e}-07} _{-3\mathrm{e}-07} $ & $4.47\mathrm{e}-06$ \\[0.1cm]
\hline   
\hline 
\end{tabular}
\tablefoot{Lower parameter limit was hit for the CGM kT parameter in both fits. CGM Abundanc and Cor norm lower parameter limits were furthermore hit in the Gammapy fit.}
\end{table*}

\FloatBarrier
\section{Exposure-weighted background}
\label{sec:app:3d_analysis}

\begin{figure}[h!]
    \centering
    \includegraphics[width=0.49\linewidth]{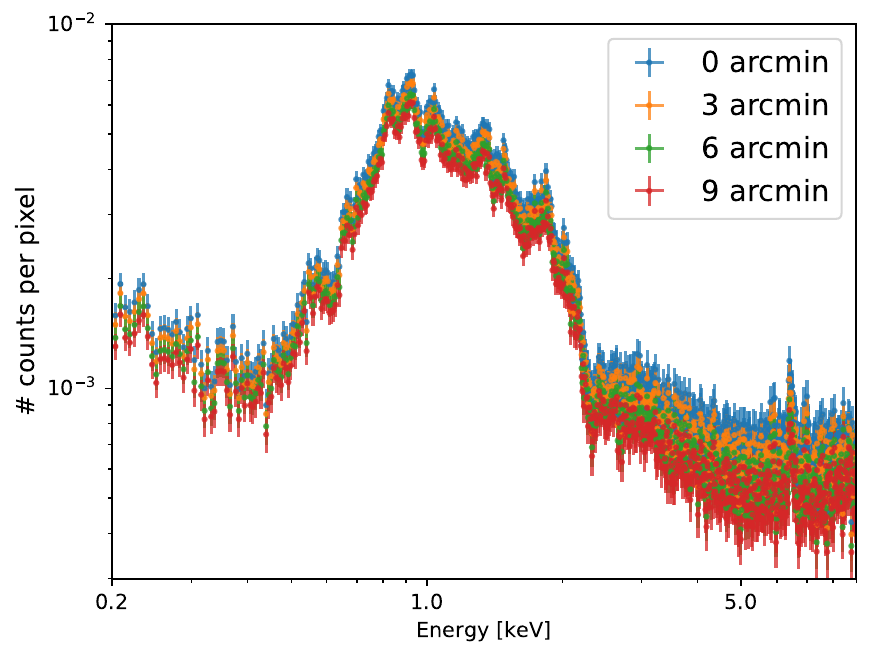}
    \caption{Comparison of EDR exposure-weighted background counts for different offsets to the pointing direction.}
    \label{fig:app:template_edr}
\end{figure}

Utilising the exposure-weighted background for pointed EDR data resulted in different background levels across the FoV, as is visualised in Fig. \ref{fig:app:template_edr}.

\end{appendix}

\end{document}